\documentclass[epj,final]{svjour}
\usepackage{stfloats}
\usepackage{graphics}
\usepackage{subfigure}
\usepackage{atlasphysics}
\usepackage{amsmath}
\usepackage{amssymb}
\usepackage{multirow}
\usepackage{lscape}
\usepackage{rotating}
\usepackage{url}
\usepackage{cite}
\usepackage{xspace}
\usepackage{verbatim}
\usepackage{subfigure}
\usepackage{hyperref}
\begin{document}

\title{\vspace{-3.0cm}\flushleft{\normalsize\normalfont{CERN-PH-EP-2011-098}} \flushright{\vspace{-0.86cm}\normalsize\normalfont{Submitted to Eur.\ Phys.\ J.\ C}} \\[3.0cm] 
\flushleft{Measurement of multi-jet cross sections in proton-proton collisions at a 7~{\bf \TeV} center-of-mass energy}}
\author{The ATLAS Collaboration}                    
\date{September 15, 2011}
\abstract{
Inclusive multi-jet production is studied in proton-proton collisions at a 
center-of-mass energy of 7~$\TeV$, using the ATLAS detector. 
The data sample corresponds to an integrated luminosity of 
2.4~pb$^{-1}$. Results 
on multi-jet cross sections are presented and compared to both 
leading-order plus parton-shower Monte Carlo predictions and to next-to-leading-order QCD calculations.
} 
\titlerunning{Multi-jet cross sections in proton-proton collisions at a 7~\TeV\ center-of-mass energy}
\maketitle
\section{Introduction}

At hadron colliders, events containing multiple jets in the final state are plentiful and provide a fertile 
testing ground for the theory of the strong interaction, quantum chromodynamics (QCD).  At high 
transverse momentum ($\pt$), the production of jets is modelled by QCD as the hard 
scattering of partons and the subsequent parton showering, followed by a hadronization process.  
Within this framework, the jet energy is related to the energy of partons produced in hadron collisions.  Consequently, the study of 
energy distributions for multi-jet events provides a fundamental and direct 
test of QCD at hadron colliders.  

In addition to their role in testing QCD, multi-jet events are often an important background in searches for new particles and new 
interactions at high energies.  
In particular, systematic uncertainties that contribute to multi-jet cross section measurements can carry over into search analyses.
Even though 
the impact of multi-jets on such analyses will vary according to 
the specific data selection criteria, a study of multi-jet events serves as an important cross check of models used to estimate backgrounds originating from jets. 

Measurements of multi-jet cross sections at the Tevatron have been performed by the CDF~\cite{Abe:1995rw, Abe:1996nn} and 
D0~\cite{Abbott:2000ua, Abazov:2002mr} collaborations in proton-antiproton collisions at 1.8 TeV center-of-mass energy. 
The CMS collaboration has recently released measurements of the three-jet to two-jet cross sections at a 7 TeV center-of-mass energy~\cite{CMS-multijets}.  
In this paper, a first study is performed of multi-jet events from proton-proton collisions at 7 TeV 
center-of-mass energy using the ATLAS detector at the Large Hadron Collider (LHC) at CERN.  
The data sample used for the analysis was collected from April until August  2010 
and represents a total integrated luminosity of 2.4~pb$^{-1}$.   
Approximately half a million events with at least two jets in the final state are selected using this data sample.  

Two primary motivations for the multi-jet study in this paper are to evaluate 
how robust leading-order perturbative QCD (LO pQCD) calculations are in representing the high jet multiplicity events, and to 
test next-to-leading-order perturbative QCD (NLO pQCD) calculations.  
For the leading-order comparisons, events with up to six jets 
in the final state are studied, and for the next-to-leading-order perturbative QCD study, the focus is on three-jet events and their comparison to two-jet events. At present, there is no four-jet NLO pQCD calculation available. 

The paper is organized as follows.  Section 2 presents a description of the ATLAS detector.   Section 3 discusses the cross sections and kinematics.  In Section 4, theoretical calculations, to which the measurements are compared, are described.  Sections 5 and 6 discuss the event selection and data corrections.  The main uncertainty coming from the jet energy scale is discussed in Section 7, followed by the results and conclusions. 

\section{The ATLAS Detector}\label{sec:detector}
The ATLAS experiment consists of an approximately 45-metre long, 25-metre diameter 
cylindrically 
shaped detector centered on the proton-proton interaction point.  A detailed 
description of the ATLAS experiment can be found elsewhere~\cite{atlas_detector_paper}. 
High-energy 
particles produced in collisions initially pass through an inner tracking system 
embedded in a 2 T solenoidal magnetic field.  The field is located in a region of 
diameter 2.3 metres and 7 metres long also centered at the interaction point. The design
of this tracking system allows the measurement of charged particle kinematics
within the pseudorapidity\footnote{ATLAS uses a right-handed coordinate system with its origin at the nominal interaction point (IP) in the centre of the detector and the $z$-axis along the beam pipe. The $x$-axis points from the IP to the centre of the LHC ring, and the $y$ axis points upward. Cylindrical coordinates $(r,\phi)$ are used in the transverse plane, $\phi$ being the azimuthal angle around the beam pipe. The pseudorapidity is defined in terms of the polar angle $\theta$ as $\eta=-\ln\tan(\theta/2)$. The rapidity is defined as $y = 0.5 \times {\rm ln}[(E+p_z)/(E-p_z)]$, where $E$ denotes the energy and $p_z$ is the component of the momentum along the beam direction. For massless objects, the rapidity and pseudorapidity are equivalent.} range of $|\eta| < 2.5$.  Precision tracking using the 
pixel detector with a space point resolution as small as 10 microns by 70 microns (in the beam 
direction) begins at a radial distance of 5 cm from the interaction point~\cite{pixel}.  The identification of the 
vertex from which the jet 
originates, performed with the inner tracker, 
is of interest in the study of multi-jet events.  

Just outside the inner tracker system are
liquid argon and scintillating tile calorimeters used for the measurement of particle 
energies.  A liquid-argon/lead electromagnetic calorimeter covers the 
pseudorapidity range of $|\eta| < 3.2$.  
This calorimeter is complemented by hadronic calorimeters, built 
using scintillating tiles and iron for $|\eta| < 1.7$ and liquid argon and copper in the 
end-cap ($1.5 < |\eta| < 3.2$).  
Forward calorimeters extend the coverage to $|\eta| = 4.9$. 
The calorimeters are the primary detectors used to reconstruct the jet energy in this analysis
and allow the reconstruction of the jet $\pt$ with a fractional resolution of better than 0.10 for
jets of $\pt=60\GeV$ and 0.05 for jets of $\pt=1\TeV$. 

Outside the calorimeters is a toroidal magnetic field that extends to the edge of the detector. 
Additional tracking detectors designed for 
measuring muon kinematics are placed within this magnetic field. 
The impact of muons in the analysis presented in this paper
is negligible.  

The ATLAS trigger system employs three trigger levels, of which only the hardware-based first level 
trigger is used in this analysis. Events are selected using the calorimeter based jet trigger. 
The first level jet trigger~\cite{atlas_trigger08} uses coarse detector information 
to identify areas in the calorimeter where energy deposits above a certain threshold occur. A 
simplified jet finding algorithm based on a sliding window of size 
$\Delta\phi\times\Delta\eta=0.8\times 0.8$ is used to identify 
these areas. This algorithm uses coarse calorimeter towers with a 
granularity of $\Delta\phi\times\Delta\eta=0.2\times 0.2$ as inputs.

\section{Cross Section Definitions and Kinematics}\label{sec:definitions}
In this analysis, the anti-$k_t$ algorithm~\cite{Cacciari_antikt,Fastjet}, with jet constituents combined according to their four-momenta, 
is used to identify jets. 
For high multiplicity studies, which includes events with 
up to six jets,
the resolution parameter in the jet reconstruction 
is fixed to $R = 0.4$ 
to contend with the limited phase space and to reduce the impact of the 
underlying event in the jet energy determination. 
For testing NLO pQCD calculations, 
where the study focuses 
on three-jet events, 
a resolution parameter of $R = 0.6$ is preferred, 
since a larger value of $R$ is found to be less sensitive to theoretical 
scale uncertainties.  The anti-$k_t$ algorithm was chosen for a variety of reasons.  It can be implemented
in the NLO pQCD calculation, is infra-red and collinear safe to all orders, and reconstructs jets 
with a simple geometrical shape. 

Jet measurements are corrected for all experimental effects such that they can be compared to particle-level predictions. 
At the particle level, jets are built using all final-state particles with a proper 
lifetime longer than 10~ps. 
These corrections
are described in Section~\ref{sec:unfolding}. The NLO pQCD calculation is 
not interfaced
to a Monte Carlo simulation with hadronization and other non-perturbative effects.  The correction for non-perturbative effects applied to the NLO pQCD calculation is described in Section~\ref{sec:theory}.

Cross sections are calculated in bins of inclusive jet multiplicity, 
meaning that an event is counted in a jet multiplicity bin if it contains a number of jets that is equal to or greater than that multiplicity.  For example, an event with three reconstructed jets will be counted both in the two-jet and three-jet multiplicity bins.
Inclusive multiplicity bins are used because they are stable
in the pQCD fixed-order calculation, unlike exclusive bins. 
Only jets with $\pt > 60\GeV$ and $|y|<2.8$ are counted in the analysis.  These cuts are chosen to ensure that the jets are reconstructed with high efficiency.  
The leading jet is further required
to have $\pt > 80\GeV$ to stabilize the NLO pQCD calculations in the dijet case~\cite{Frixione}. 

\section{Theoretical Predictions}\label{sec:theory}
Measurements are compared to pQCD calculations at leading order and next-to-leading order. 

Many different effects are included in leading-order Monte Carlo simulations of jets at the LHC. 
These include the modeling of the underlying event and hadronization, which can affect the
cross section calculation through their impact on the jet kinematics~\cite{UE-ATLAS}.  
Effects arising from differences between the matrix-element plus parton-shower 
(ME+PS) calculation (with up to $2\rightarrow n$ matrix-element scattering diagrams) and 
the parton-shower calculation alone 
(with only $2\rightarrow 2$ matrix-element scattering diagrams) also need to be understood. 
These topics are not easily separable, since tuning of some of the
effects (such as the underlying event) to data is needed, and the tuning process 
fixes other
inputs in the Monte Carlo simulation, such as the proton parton distribution functions (PDF), the parton-shower model, and 
the hadronization model.
The inability to separate out some effects makes it difficult to obtain a full estimate
of the theoretical uncertainty associated with the leading-order Monte Carlo predictions.  Furthermore, leading-order Monte Carlo predictions are affected by large normalization uncertainties.  

In this study, the goal is to test the performance of the different leading-order Monte Carlo simulations, so that they can be used to estimate multi-jet backgrounds for new particle searches, not to discern whether deviations with respect to QCD are present in the data.  The latter goal is best achieved by comparing with NLO pQCD calculations (discussed later in this section).  For these reasons, the leading-order Monte Carlo predictions are all normalized to the measured inclusive two-jet cross section and then used for shape comparisons.  No attempt is made to assign a theoretical uncertainty to these leading-order predictions.  Instead,
numerous different Monte Carlo simulations and currently available tunes have been studied 
in order to investigate the impact of each of these effects on the measurements. Only 
a representative subset is shown in the results, even though conclusions are drawn 
on the basis of all simulations studied. 

For the leading-order analysis, ALPGEN~\cite{Mangano:2002ea} is used to 
generate events with up to six partons in the 
final state using the leading-order set of proton PDFs
CTEQ6L1~\cite{Pumplin:2002vw}. 
A factorization and renormalization scale, $Q$, that varies from event to event is used in the
event generation, where $Q^2=\sum \pt^2$.
The sum runs over all final state partons. 
ALPGEN is interfaced to 
PYTHIA 6.421~\cite{Sjostrand:2007gs,Sjostrand:2000wi} and, alternatively, to
HERWIG/JIMMY \cite{Corcella:2000bw,Corcella:2002jc,Butterworth:1993ig,Butterworth:1996zw} to 
sum leading logarithms to all orders in the parton-shower approximation and to include 
non-perturbative effects such as hadronization and the underlying event.
The ATLAS generator tunes from 2009 (MC09$^{\prime}$\footnote{
The ATLAS MC09$^{\prime}$ tune only differs from MC09 tune in the value of one 
parameter regulating multiple interactions, 
PARP(82), which is the same used in the MC08 tune\cite{ATL-PHYS-PUB-2010-002}.}
~\cite{ATL-PHYS-PUB-2010-002}) and from 2010 (AUET1\cite{ATL-PHYS-PUB-2010-014}) 
are used. 
Additional tunes have been 
investigated to assess the impact of the underlying-event and parton-shower tuning. 
With comparable underlying-event tunes and ALPGEN
parameters, the comparison between ALPGEN+PYTHIA and ALPGEN+HERWIG/ JIMMY uncovers 
differences that may arise from different parton-shower implementations and 
hadronization models.

SHERPA~\cite{Sherpa} with its default parameters and renormalization scale scheme from 
version 1.2.3 is also used to generate events with up to six partons in the final state.  This provides 
an independent matrix-element calculation with a different matching scheme between the 
matrix element and the parton shower.  
Detailed studies of individual tunes using SHERPA, however, are not performed in this paper.

The PYTHIA 
event generator is also compared
to the data to study the limitations of leading-order $2\rightarrow 2$ matrix-element calculations. 
This generator implements a leading-order 
matrix-element calculation for $2\rightarrow2$ processes, $\pt$-ordered parton showers, 
an underlying-event 
model for multiple-parton interactions and the Lund string model for hadronization. The 
MRST2007 modified leading order~\cite{Sherstnev:2008dm,Martin:2009iq} PDFs interfaced with the AMBT1\cite{ATL-PHYS-PUB-2010-002} 
generator tune are used in the sample generation. 

For the purpose of understanding detector effects, the particles generated in the leading-order
Monte Carlo generators are passed through a full 
simulation of the ATLAS detector and trigger~\cite{SimJINST} based on GEANT4~\cite{geant}.  
Additional proton-proton collisions are added to the hard scatter in the simulation process
to reproduce realistic LHC running conditions. 
Events and jets are selected using the same criteria in data and Monte Carlo simulations. 

For the next-to-leading-order pQCD study, the calculation implemented in NLOJet++ 4.1.2~\cite{NLOJet} 
is used. 
The renormalization and factorization scales are varied independently by a factor of two in order to estimate the impact of higher order terms not included in the calculation.  An additional requirement that the ratio of the renormalization and factorization scales did not differ by more than a factor of two was imposed.
Two next-to-leading-order PDF sets, 
CTEQ 6.6~\cite{Nadolsky:2008zw} and MSTW 2008 NLO~\cite{Martin:2009iq}, 
are used for calculating the central values. Only results obtained with the MSTW 2008 NLO PDF 
set are shown in the paper since the results obtained with the CTEQ 6.6 PDF set are compatible. 
The 90\% confidence-limit
error sets are used in the evaluation of the PDF uncertainties.
The uncertainty
in the calculations due to the uncertainty in the value of $\alpha_S$ is determined by varying the value of 
$\alpha_S$ by $\pm0.002$ for each PDF set. 

The NLOJet++ program implements a matrix-element calculation, and therefore it lacks a 
parton-shower interface and
does not account for non-perturbative effects. To compare to particle-level
measurements, a correction factor is required. PYTHIA and 
HERWIG++~\cite{Herwigpp}
are used to generate samples without underlying event. Jets in these samples are reconstructed
from partons after the parton shower, and observables are compared to those obtained 
at the particle level in the standard HERWIG++ and PYTHIA samples. A 
multiplicative correction is calculated 
\begin{equation}
C_{\rm non-pert}=\frac{o^{\rm particle}_{\rm UE}}{o^{\rm parton}_{\rm no\; UE}},
\end{equation}
where $o$ is the observable of interest calculated at the particle or parton level in the samples
with and without underlying event. 
The correction factor takes the next-to-leading-order pQCD calculations to the particle level. 
This correction is calculated in 
three different samples. The correction
obtained using the PYTHIA AMBT1 sample is taken as the default 
value for the analysis, and the 
systematic uncertainty is estimated from the
maximum spread compared to the results from the other models (marked with an
asterisk in Table~\ref{tab:samples}). The size of this
correction is less than 5\% in all observables studied in the next-to-leading-order pQCD analysis. 
The total uncertainty quoted on the next-to-leading-order pQCD calculations comes from the 
quadrature sum of the uncertainties from the renormalization and factorization scales, the proton PDFs, $\alpha_S$ 
and the non-perturbative corrections. 

Table~\ref{tab:samples} presents a summary of the different Monte Carlo generators and tunes 
that the data are compared to in this paper. 
\begin{table*}[htbp]
  \begin{center}
    \begin{tabular}{|c|c|c|}\hline
	Generator & PDF & tune  \\
	\hline
	ALPGEN+HERWIG/JIMMY & CTEQ6L1~\cite{Pumplin:2002vw} & AUET1\cite{ATL-PHYS-PUB-2010-014}  \\ 
         ALPGEN+PYTHIA & CTEQ6L1~\cite{Pumplin:2002vw} & MC09$^{\prime}$~\cite{ATL-PHYS-PUB-2010-002} \\ 
         PYTHIA & MRST2007 LOmod~\cite{Sherstnev:2008dm,Martin:2009iq} & AMBT1\cite{ATL-PHYS-PUB-2010-002} \\ 
	PYTHIA$^{*}$  & MRST2007 LOmod~\cite{Sherstnev:2008dm,Martin:2009iq}  & MC09~\cite{ATL-PHYS-PUB-2010-002} \\ 
	SHERPA & CTEQ66~\cite{Nadolsky:2008zw} & Default (v1.2.3) \\ 
	HERWIG++$^{*}$ & MRSTMC al~\cite{Sherstnev:2008dm,Martin:2009iq} & Default (v2.5) \\ 
		\hline
    \end{tabular}
    \caption[Samples]{
	Different Monte Carlo generators and tunes 
	used for the leading-order analysis in this paper. The asterisk indicates the samples
	used to determine the uncertainties on the non-perturbative correction to the 
	next-to-leading-order pQCD calculations.
      \label{tab:samples}}
  \end{center}
\end{table*}

\section{Event Selection and Reconstruction}\label{sec:event-sel}

\subsection{Trigger Selection}\label{sec:trigger-sel}
A set of ATLAS first level (level-1) multi-jet triggers is used to select events for the analysis. 
Multi-jet triggers require several jets reconstructed with a level-1 sliding window algorithm.  All multi-jet triggers are symmetric, meaning that each trigger had one particular transverse energy threshold and that this threshold was the same for all jets in an event.
Only two-jet and three-jet triggers were needed for the
analysis.  

The single-jet triggers with
a $10\GeV$ level-1 threshold
have been shown to be fully efficient for events with at least one anti-$k_t$ jet with $R=0.4$ 
and calibrated $\pt > 60\GeV$~\cite{inclusive} using events triggered with the minimum bias
triggers. 
The efficiency for triggering on the leading jet
is calculated using the minimum bias triggers. 
Then, the efficiency of the trigger to fire on
the second leading jet is calculated by requiring that the leading jet passes the 
single-jet trigger.  Similarly, the efficiency of the third leading jet is studied by requiring that the second leading jet is matched to a jet trigger object, and the event passes a two-jet trigger.  For $\pt > 60 \GeV$, events are selected on the trigger plateau. 

\begin{figure*}[htbp]
 \centering
 \subfigure[]{
   \includegraphics[width=0.4\textwidth]{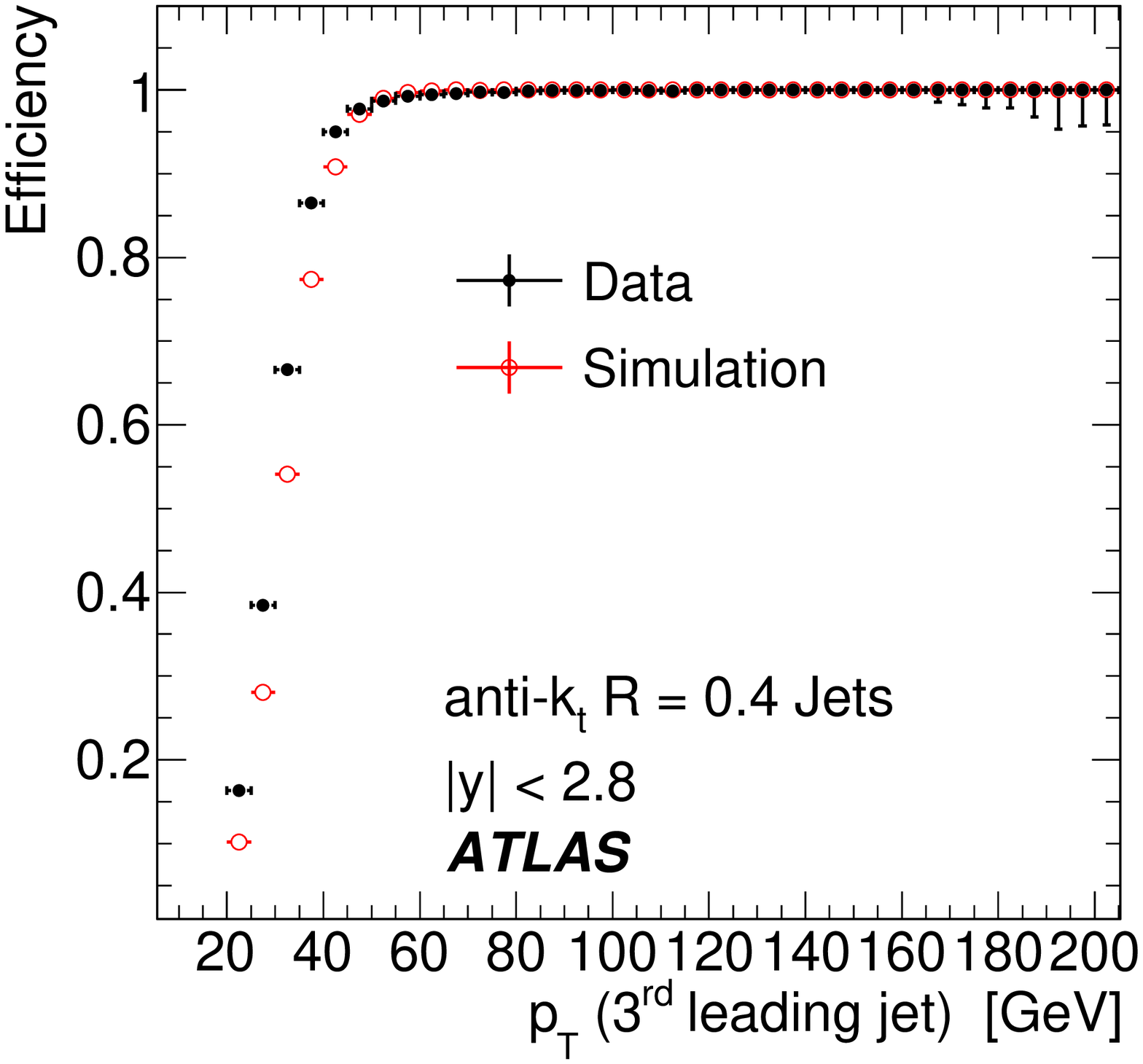}
}
 \subfigure[]{
    \includegraphics[width=0.4\textwidth]{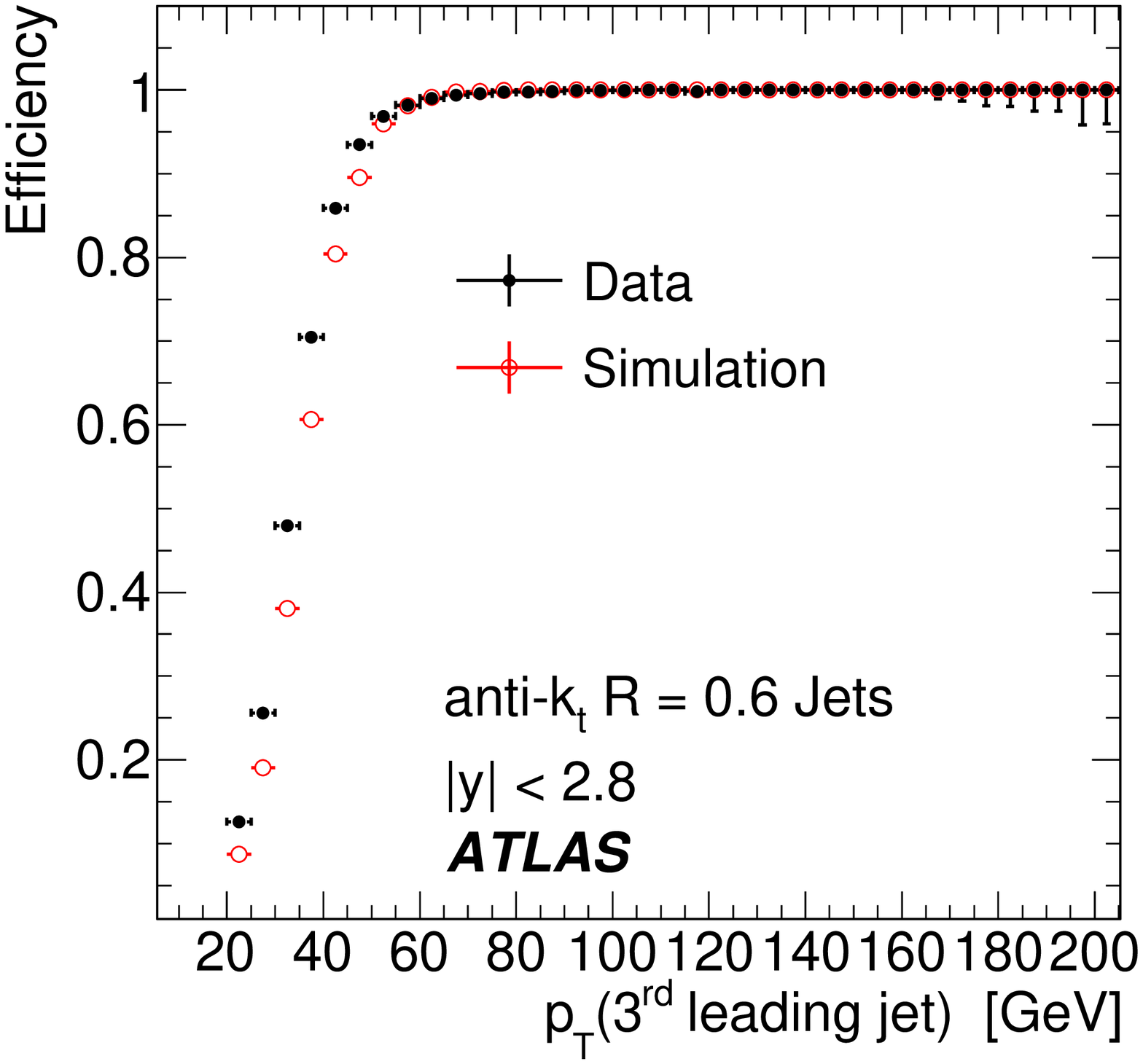}   
}
 \subfigure[]{
   \includegraphics[width=0.4\textwidth]{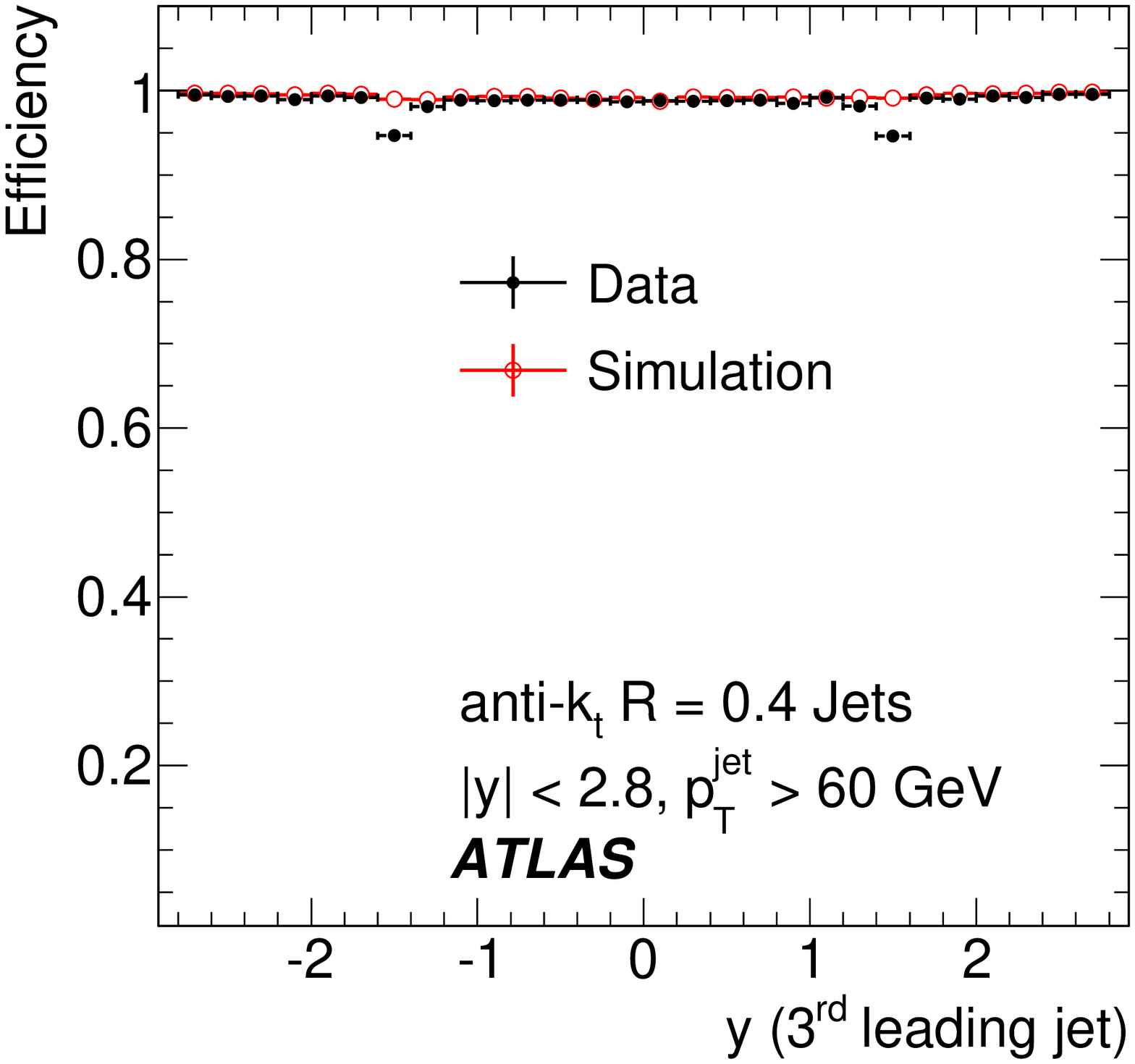}
}
 \subfigure[]{
    \includegraphics[width=0.4\textwidth]{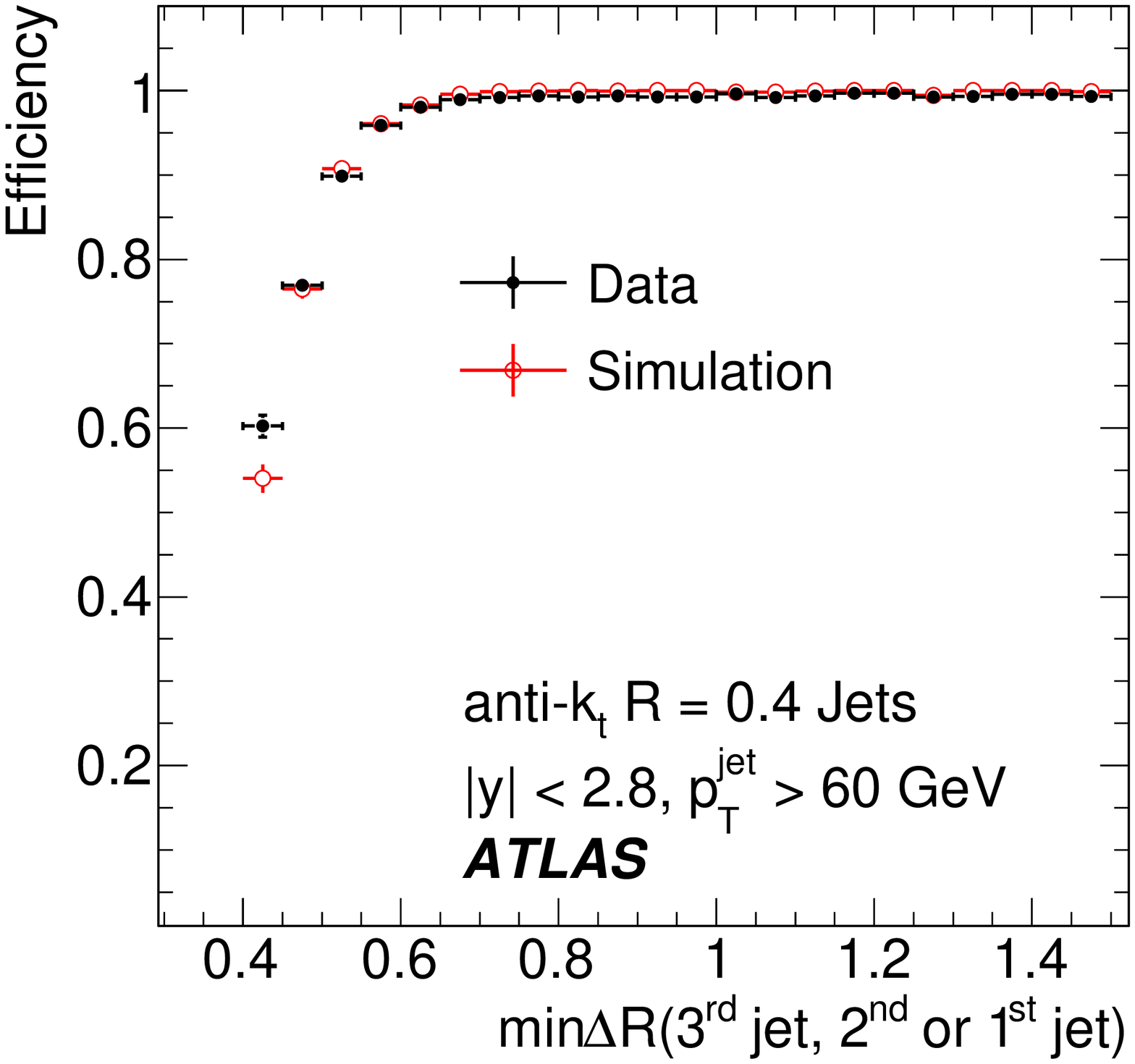}   
}
\caption{Jet trigger efficiency for the third leading jet as a function of $\pt$ for anti-$k_t$ jets with  $R=0.4$ (a), and $R=0.6$ (b). Jet trigger efficiency as a function of $y$ of the third leading jet
 with $\pt>60\GeV$ and $R=0.4$ (c). 
 Jet trigger efficiency as a function of the minimum separation $\Delta R$ between the two closest jets (d).
      The efficiency is shown both as calculated in data, as described in the 
      text, and in Monte Carlo simulations for the three-jet trigger with a level-1 cut on the jet transverse energy 
      of $10\GeV$. 
    }
   \label{fig:trigger6}
\end{figure*}

Figure~\ref{fig:trigger6} shows the efficiency
for the third leading jet to fire the three-jet trigger as
a function of the reconstructed jet $\pt$ for jets of $R=0.4$ (a) and $R=0.6$ (b). 
The efficiencies calculated in data are 
compared to those from the Monte Carlo detector simulation. 
The efficiency as a function of jet rapidity is also shown for $R=0.4$ jets (c) for
$\pt > 60\GeV$.  A small inefficiency is present in the data at $y = \pm 1.5$.
In this transition region between the barrel and end-cap calorimeters the level-1 trigger 
energy sums did not span between the calorimeters for the early data used here, 
resulting in this small efficiency drop, which is not modelled by the Monte Carlo simulation. 
The simulation is not corrected for this effect, since its impact in the measurements is negligible, 
and included as part of the systematic uncertainties in the data correction described in 
Section~\ref{sec:unfolding}. 

The event-level efficiency as a function
of the closest distance between two selected $R=0.4$ offline jets for events selected using the 
three-jet
trigger is shown in Figure~\ref{fig:trigger6} (d).  The study probes possible topological dependences in the trigger. A
dependence at low 
$\Delta R$ 
is observed, where
$\Delta R=\sqrt{\Delta\phi^2+\Delta \eta^2}$ represents the minimum separation between selected jets in the event.
The dependence on 
$\Delta R$ 
 is well described by the Monte Carlo simulation. 
For the calculation of the efficiency in the data, the
two leading jets are associated with level-1 jet objects and 
an assumption is made that any topological inefficiency will only affect one of the level-1 jet 
objects. 
Figure~\ref{fig:trigger6} (d) indicates that events in which two jets are separated by $\Delta R < 0.6$ have an efficiency of less than 100\%.
This inefficiency
appears to depend weakly on the jet $\pt$ and is well
described in the detector simulation for events where the closest distance between
selected jets is greater than 0.45. 
The inefficiency is accounted for in the Monte Carlo-based data correction
described in 
Section~\ref{sec:unfolding}.  Such an inefficiency is not observed in the analysis of jets reconstructed using the anti-$k_t$ algorithm with resolution parameter $R = 0.6$.

The three-jet trigger operated without
pre-scaling for the entire data collection period used in this paper. 
All events falling in the three-jet inclusive multiplicity bin are, therefore, selected using the three-jet trigger with a jet threshold of $10\GeV$ on the level-1 jet objects.  
On the other hand, a 
large pre-scaling was applied to
certain two-jet triggers.  In order to select events in the two-jet {\em inclusive} multiplicity bin,
several two-jet triggers were used.
Three two-jet triggers with symmetric transverse energy thresholds of 10, 15 and $30\GeV$ were combined independently, weighted by the integrated luminosity associated with each trigger.  
The three triggers were combined in such a way that only one of them was
responsible for counting events
for which the $\pt$ of the second leading jet was in a particular range.
 Specifically, the three triggers with thresholds of 10, 15 and 30$\GeV$ covered the ranges of second leading jet $\pt$ of 60-80$\GeV$, 80-110$\GeV$ and greater than 110$\GeV$, respectively. The two-jet triggers have an efficiency higher than 99\% to select such events. 

\subsection{Vertex Reconstruction}
The primary vertex or vertices are found using tracks that originate near the beam collision 
spot\cite{ATLAS-CONF-2010-027}, satisfy quality criteria~\cite{ATLAS-CONF-2010-069} 
and have transverse momentum above $150\MeV$. A vertex is seeded by searching for 
the global maximum in the distribution of $z$ coordinates of reconstructed tracks. 
The vertex is fitted using the position of this seed along with neighboring tracks. Tracks 
incompatible with the reconstructed vertex are used to seed new vertices until no tracks are left. 
This analysis only uses events
in which at least one primary vertex with at least five associated tracks has been reconstructed. 
No cut on the primary vertex position is applied. The {\em event} vertex 
is defined as the vertex in the event for which the sum of the 
$\pt$ of the tracks associated to that vertex is largest. 

\subsection{Jet Reconstruction}
Topological clusters of calorimeter energy evaluated at the electromagnetic scale~\cite{inclusive} 
are used as inputs to the jet finding algorithm. These clusters use the baseline calibration derived
from test beams and from $Z\rightarrow ee$ data~\cite{atlas_Z_paper}, 
which reconstructs the energy of particles interacting electromagnetically. 
The anti-$k_t$ algorithm~\cite{Cacciari_antikt} with resolution parameters 
$R=0.4$ and $R=0.6$ and full four-momentum recombination 
is used to reconstruct jets from clusters. The jet 
four-momentum is calculated assuming that the jet 
origin is at the position of the event vertex. 
The jet reconstruction is fully efficient in the Monte Carlo 
simulation for jets with transverse momentum above $30\GeV$. 
The reconstruction efficiency in the simulation compares well with the one 
measured with data~\cite{inclusive}. 

\subsection{Jet Energy Scale Calibration}
Jets reconstructed at the electromagnetic scale are measured to have an energy which 
is lower than the true energy of interacting particles within the jet. 
The difference between a hadron-level jet and an electromagnetic-scale jet is due to the different calorimeter response to electromagnetic objects compared to strongly interacting objects, detector induced showering and energy deposition in regions of the detector that are not instrumented. 
A Monte Carlo-based calibration 
that corrects for these effects as a function of $\pt$ and $y$ is used to obtain 
jets with the correct energy scale\cite{atlas_jes_study}.

\subsection{Jet Selection Criteria}\label{sec:offline-sel}
Jets considered in the analysis are selected using the following kinematic and data quality
selection criteria:
\begin{enumerate}
\item The event must contain at least one jet with $|y| < 2.8$ and a $\pt$ greater than 
$80\GeV$. 
\item Jets are required to have $|y| < 2.8$ and $\pt > 60\GeV$ in order to be counted.
\item A series of jet cleaning cuts were applied to eliminate various detector effects and suppress
beam and other non-collision backgrounds. 
Overall, these cuts reduce the total number of jets by less than 0.1\%.
These cuts have 
been shown to be efficient in eliminating
noise, while rejecting a negligible number of true jets. 
\item In order the reduce the effects from pileup events, jets are only accepted if at least 70\%  
of their charged 
particle $\pt$ comes from the event vertex. The charged particle $\pt$ is calculated as the
scalar sum of the $\pt$ of reconstructed tracks within a $\Delta R$ equal to the resolution
parameter used in the jet reconstruction. 
Overall, this cut lowers the number of selected two-jet events  
by 0.4\%, and its effect increases with jet multiplicity.  The cut reduces the number of selected 
six-jet events by 
3.4\%. 
All observables show a
negligible dependence on the number of reconstructed primary vertices once this cut is applied\cite{njet2011}. 
Jets with no charged particle content are accepted, but only constitute a few percent of events at low $\pt$.
\item Only events with at least two selected jets are used in the analysis.
\end{enumerate}

\begin{figure*}[htbp]
 \centering
   \includegraphics[width=0.9\textwidth]{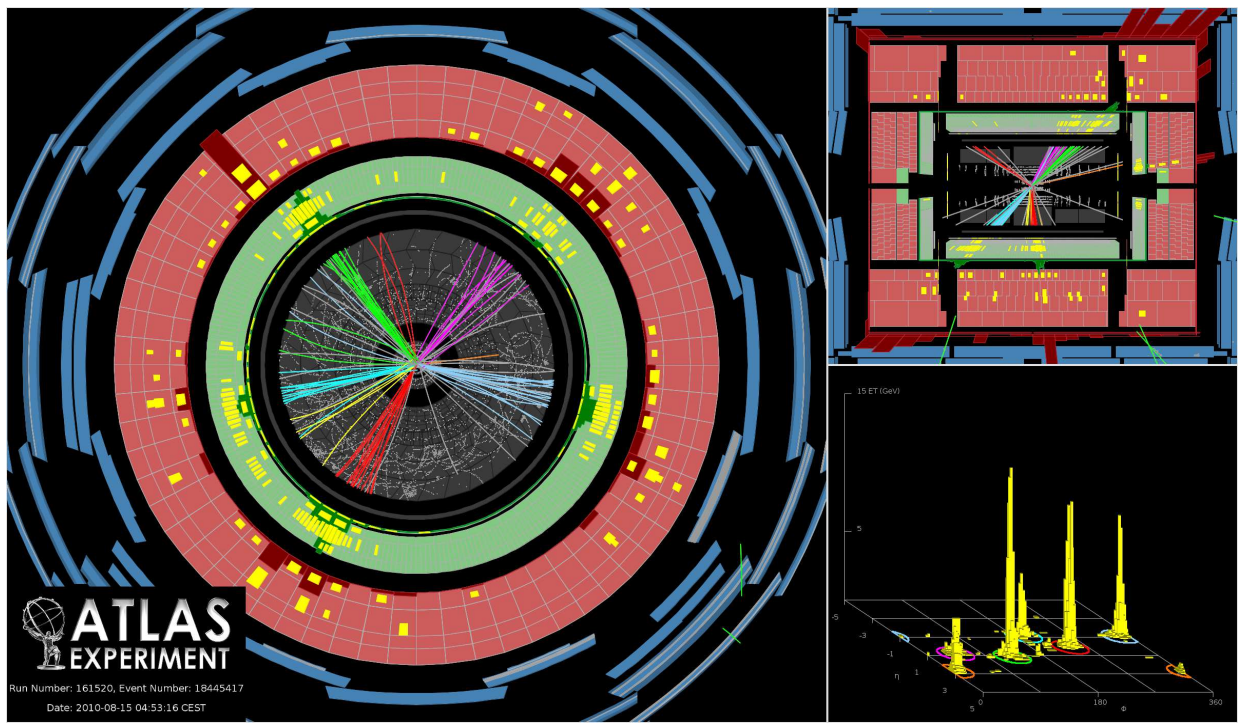}
    \caption{ Event display of a six-jet event satisfying the analysis requirements.  
    The towers in the bottom right figure represent transverse energy 
    deposited in the calorimeter projected 
    on a grid of $\eta$ and $\phi$. Jets with transverse momenta ranging from 84 to $203\GeV$ 
    are measured in this event.}
   \label{fig:event-display}
\end{figure*}

For illustrative purposes, Figure~\ref{fig:event-display} presents an event display of a 
six-jet event passing all selection cuts. The transverse energy
deposition in the calorimeter is shown as a function of $\eta$ and $\phi$.  
For this event, the six selected jets are well separated spatially.

Table~\ref{tab:nevents} presents the total number of multi-jet events 
versus inclusive jet multiplicity. No correction for trigger pre-scales in the two-jet bin 
has been applied to the numbers in the table. 
\begin{table}[htbp]
  \begin{center}
    \begin{tabular}{|c|c|}\hline
	Inclusive multiplicity & Number of events \\
	\hline
	$\ge$ 2 & 500,148 \\
	$\ge$ 3 & 112,740  \\
	$\ge$ 4 & 10,999  \\
	$\ge$ 5 & 1,100 \\
	$\ge$ 6 & 115  \\
	\hline
    \end{tabular}
    \caption[Selected events]{
	Number of selected events using the criteria described in this paper as a function of 
	inclusive jet multiplicity for jets reconstructed with the anti-$k_t$ algorithm with resolution parameter $R=0.4$ before correcting for trigger pre-scales.  
      \label{tab:nevents}}
  \end{center}
\end{table}

\section{Data Correction for Efficiencies and Resolution}\label{sec:unfolding}
A correction is needed to compare the measurements 
to theoretical predictions. 
The correction, which accounts for 
trigger inefficiencies, detector resolutions and other detector effects that affect the
jet counting, is performed in a single step using a bin-by-bin
multiplicative factor   
calculated from Monte Carlo simulations. For
each measured distribution, the corresponding Monte Carlo simulation cross
section using truth jets as defined in Section~\ref{sec:definitions} is
evaluated in the relevant bins, along with the equivalent
distributions obtained after the application of detector simulation and analysis
cuts. The ratio of the true to the simulated distributions
provides the multiplicative correction factor to be applied to the 
measured distributions. The bins are chosen so that bin migrations due to resolution effects
are small. Typically, above 70\% of events in a bin built using reconstructed quantities come
from the same bin using particle-level quantities in the simulation. 
A similar fraction of events in a given truth bin
fall in the same bin using reconstructed quantities. 
These fractions, which characterize bin migrations, 
become smaller with increasing jet multiplicity, but never become less than 0.6. 

To perform the correction, the 
ALPGEN+HERWIG/ JIMMY AUET1 Monte Carlo simulation is used. The sample includes, 
on average,  
two additional soft proton-proton collision events overlapping with the hard scatter simulated
by ALPGEN. 
The data have fewer overlapping collisions, as revealed by the distribution of the number
of selected vertices, and the 
Monte Carlo simulation is subsequently weighted to match the distribution from the data. The truth distribution is independent
of the additional collisions, since jets are built using particles simulated by the 
ALPGEN+HERWIG/JIMMY Monte Carlo simulation only. Distributions in the Monte Carlo
simulation are not further reweighted to match the data. The impact of differences in 
shapes between data and Monte Carlo simulation on the
calculation of the correction factors is instead considered 
part of the systematic uncertainties in these factors.

\begin{figure*}[htbp]
 \centering
 \subfigure[]{
   \includegraphics[width=0.45\textwidth]{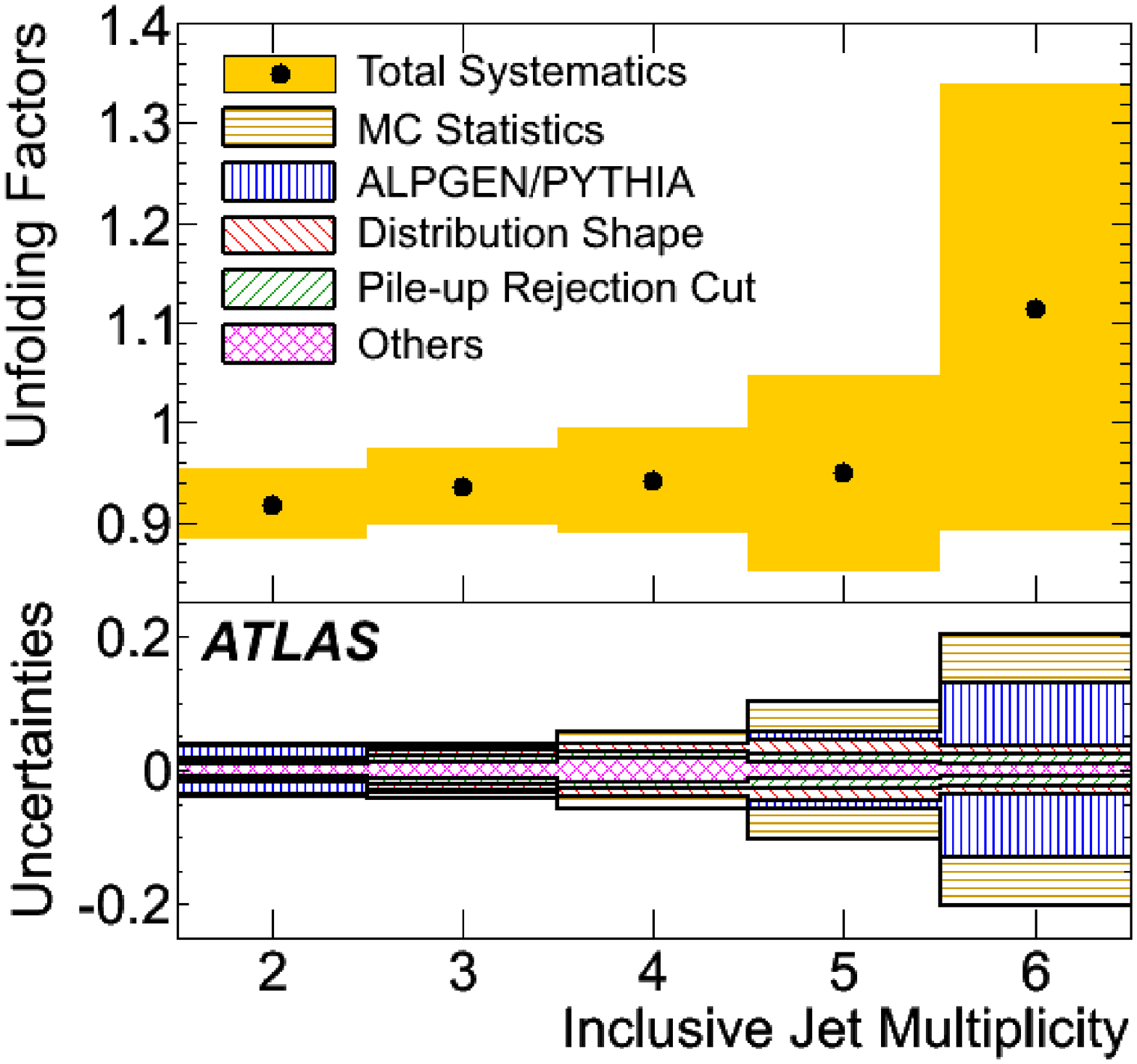}
}
 \subfigure[]{
    \includegraphics[width=0.45\textwidth]{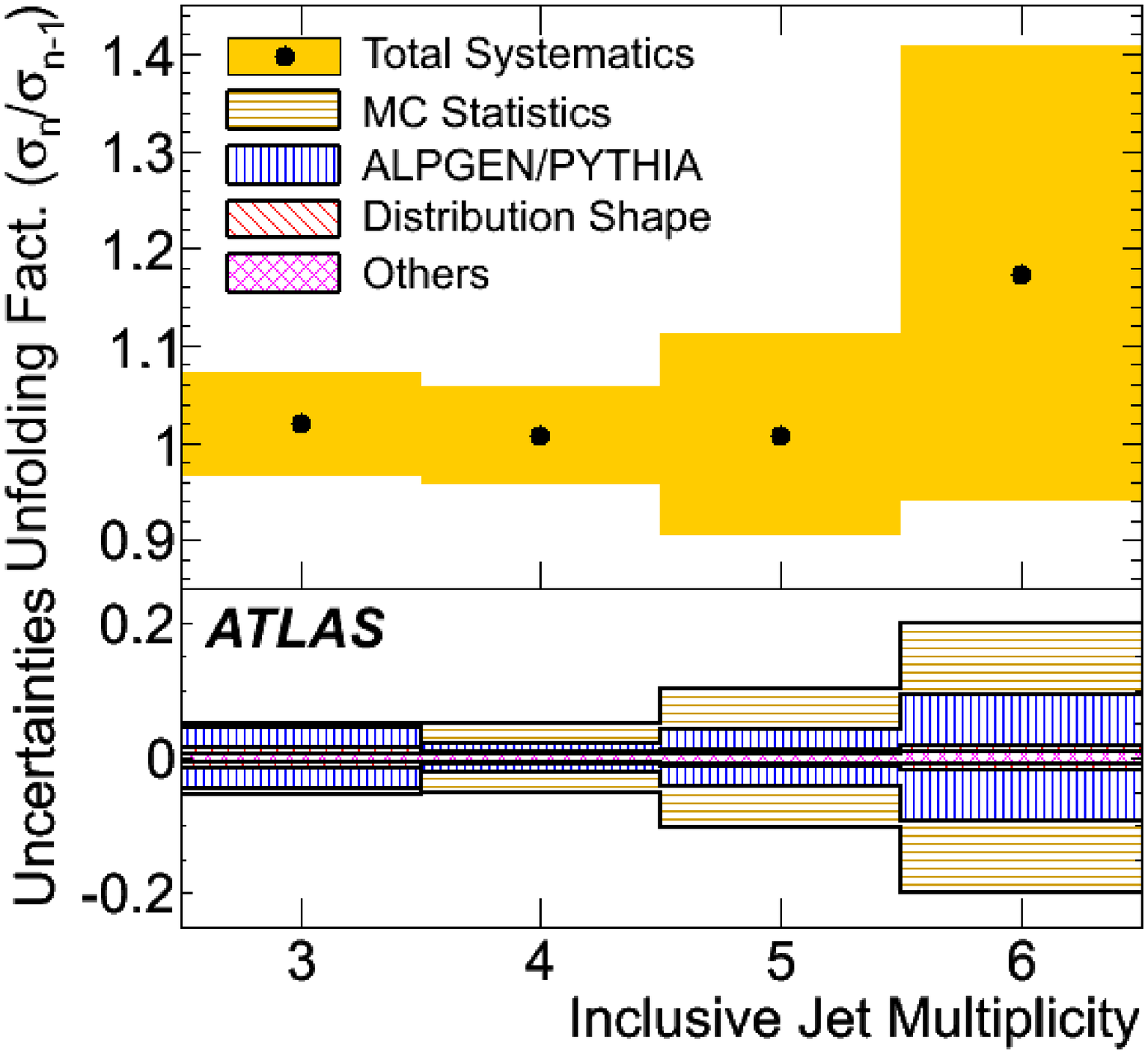}   
}
    \caption{Bin-by-bin correction factors for the cross sections (a) 
    and for the $n$ to $n-1$ cross-section ratios (b) as a function of the inclusive jet multiplicity. The 
    correction factors calculated using the ALPGEN+HERWIG/JIMMY AUET1 sample are shown with the systematic 
    uncertainty as a yellow band around the points. See the text for an explanation of the legend labels. }
   \label{fig:unfoldingFactors}
\end{figure*}

The uncertainty in the correction factors is estimated taking into account several effects. 
One arises from the spread in correction factors coming from different generators
(ALPGEN+ HERWIG/JIMMY AUET1 and PYTHIA AMBT1).  A second detailed study is performed in which the simulated jet $\pt$, $y$ and $\phi$ resolution is varied according to their measured
uncertainties~\cite{ATLAS-CONF-2010-054,daCosta:2011ni}. Third, the shape of the simulated distributions is varied within limits set by the present measurements 
in order to account for possible biases 
caused by the input distributions. Samples with a trigger inefficiency in the crack
region, with different pile-up rejection cuts and different primary vertex multiplicity distributions are also used to 
estimate the uncertainty arising from trigger effects and from the impact of overlapping proton-proton
collisions.
All these effects impact the systematic uncertainties in the correction factors, 
and their uncertainties are ultimately added in 
quadrature to provide the final systematic uncertainty in the bin-by-bin correction. Although only important for particular bins, statistical
uncertainties on the correction factors are
added to the total uncertainty.
Results for the bin-by-bin correction factors are presented in Figure~\ref{fig:unfoldingFactors}. The corresponding
uncertainties are calculated for the cross section (a) and for the $n$ to $n-1$ cross-section ratios (b) 
as a function of the inclusive jet multiplicity. 
The combined systematic uncertainty is shown as a yellow band around the correction factors.
The main components contributing to the systematic uncertainty are shown 
at the bottom of each figure. 
The uncertainty in the correction factors for detector efficiencies and resolutions is
smaller for most bins and observables than the uncertainty coming from the jet energy scale calibration, discussed in the next section.

The systematic uncertainties in the luminosity calculation affect all cross section measurements,
but cancel out in all measurements where cross-section ratios are involved. The integrated luminosity
of the dataset used in this paper is measured to be 2.43$\pm$0.08~pb$^{-1}$~\cite{lumi}
and the associated uncertainty is not shown in the figures. 

\section{Uncertainty on the Jet Energy Scale}
The jet energy scale uncertainty is the dominant uncertainty for
most results presented in this paper. The fact that cross sections fall steeply  
as a function of jet
$\pt$ implies that even a relatively small uncertainty in the determination of the 
jet $\pt$ translates into a 
substantial change in the cross sections as events migrate along the steeply falling curve. 

The jet energy scale and its uncertainty\cite{atlas_jes_study} have been determined 
for jets from a dijet sample without nearby activity in the calorimeter.  
For a multi-jet analysis, additional systematic uncertainties need to be considered.
These uncertainties arise from the difference in the calori- meter response to jets of different flavors
as well as the impact of the presence of nearby activity in the calorimeter on the jet energy 
measurement.

\begin{figure}[htbp]
 \centering
   \includegraphics[width=0.45\textwidth]{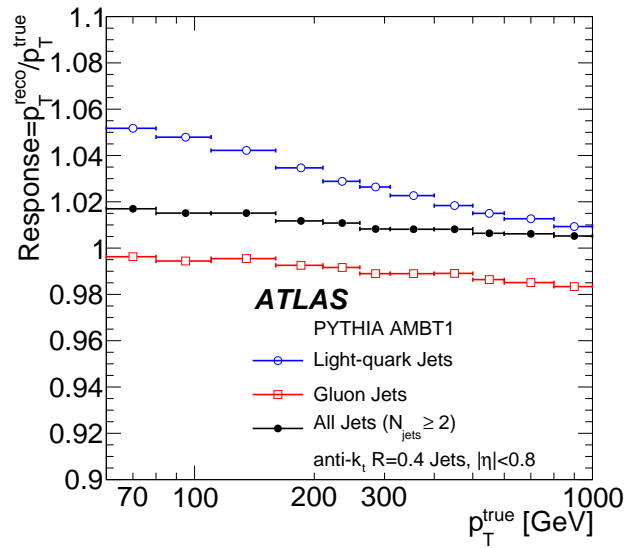}
    \caption{Jet response (mean reconstructed jet $\pt$ over true jet $\pt$) as a function of the 
    true $\pt$ for jets tagged as originating from a light quark or a gluon.  The jet 
    response in a sample with at least two jets of $\pt > 60\GeV$ (and with those two jets within
     $|y|<2.8$) 
    is also shown for those jets with $|\eta|<0.8$. 
    The anti-$k_t$ algorithm with $R=0.4$ is used.}
   \label{fig:lightGluonResponse}
\end{figure}

Figure~\ref{fig:lightGluonResponse} shows the calorimeter $\pt$ response for light-quark 
and gluon jets in the region $|\eta| < 0.8$ as a function of the true jet $\pt$ calculated using the 
PYTHIA AMBT1 Monte Carlo simulation sample. The response for jets in the two-jet inclusive multiplicity bin 
is also shown. Light-quark and gluon jets were 
tagged using the highest-energy parton found in the Monte Carlo simulation particle record
within a cone of radius equal to the
resolution parameter of the jet algorithm. Only jets that had no additional reconstructed jet of 
$\pt>7\GeV$ evaluated at the electromagnetic scale within $\Delta R=1.0$ 
from the jet axis 
were used in order to decouple effects in the response caused by jet flavor from effects 
related to the presence of nearby calorimeter activity. 

The Monte Carlo simulation shows a slightly higher fraction of jets matched to gluons for high-multiplicity
final states, particularly in the ALPGEN samples. 
To the extent that the Monte Carlo simulation reflects the data, the difference in response
as a function of multiplicity is accounted
for in the bin-by-bin correction for efficiencies and resolution. 

An additional jet energy scale uncertainty, however, could arise, since the standard jet energy 
scale was derived for a particular admixture of 
light-quark and gluon jets.  For a different admixture, the jet energy scale uncertainty could be different. In what
follows, this uncertainty is referred to as the `flavor response' uncertainty. 
This uncertainty is estimated using Monte Carlo simulations~\cite{atlas_jes_study} 
by studying the difference between the gluon and light-quark 
jet response under various assumptions. However, the relative change of the light-quark jet response
with respect to the gluon jet response is found to be negligible in all simulations 
studied~\cite{ATLAS-CONF-2011-053}, so the effect can be safely ignored. 

In addition, the fraction of light-quark and gluon jets in multi-jet samples in the data could differ 
from the fraction predicted by the Monte Carlo simulations, thus leading to a systematic shift in the 
jet energy scale. The precision with which the flavor composition of the sample is known thus
also affects the precision of the jet energy measurement. 
The flavor composition depends on 
many theoretical aspects in the event production (parton distribution functions, limitations of
leading-order calculations, initial and final state radiation tuning) and the uncertainty in the
predictions is not easy to estimate using Monte Carlo simulations. 
The uncertainty is determined using a data-driven method that provides a 
measurement of the flavor composition up to the four-jet inclusive multiplicity 
bin and for jets of $\pt<210\GeV$~\cite{ATLAS-CONF-2011-053}.  The method uses 
template fits to the distribution of jet widths and to the number of tracks associated with jets
in bins of $\eta$, $\pt$, jet isolation and jet multiplicity. 
The templates are obtained using
Monte Carlo simulations modified to match the distributions found in the two-jet bin. 
Using these template fits, the measurement of the flavor composition is determined to an accuracy of $\approx 10\%$.
Overall, ALPGEN predicts the correct flavor composition to within 30\% in bins where
the number of collected events is enough to perform the fits. 
At high $\pt$ and high multiplicities the flavor composition is assumed to be unknown when 
calculating the jet energy scale uncertainty. 

\begin{figure}[htbp]
 \centering
   \includegraphics[width=0.45\textwidth]{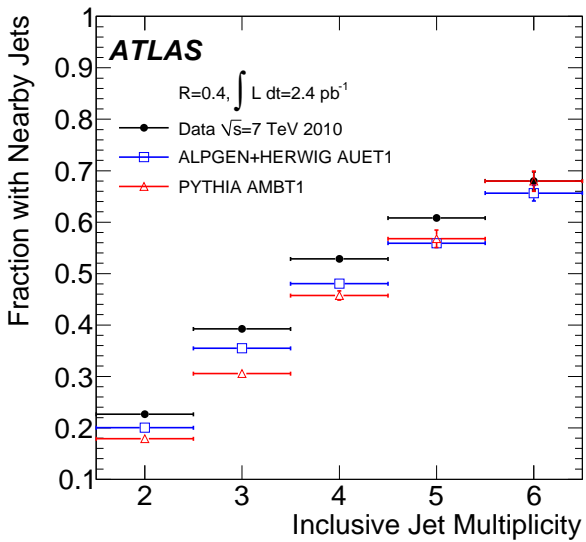}
    \caption{Fraction of selected jets in each inclusive multiplicity bin with neighboring 
    jets within $\Delta R= 1.0$.  Data (solid circles) are 
    compared to the ALPGEN+HERWIG/JIMMY AUET1 (open squares) and PYTHIA AMBT1 (open triangles) Monte Carlo simulations.
}
\label{fig:nearbyJets}
\end{figure}

Jets with nearby activity have different properties than the jets used to estimate the jet energy scale uncertainty.  
In addition, the 
fraction of jets with nearby activity increases with jet multiplicity.
Figure~\ref{fig:nearbyJets} gives the probability of a selected jet occurring within $\Delta R= 1.0$ of a reconstructed jet with 
$\pt>7\GeV$ at the electromagnetic scale as a function of inclusive 
jet multiplicity. 
The overlap probability increases with jet multiplicity, 
a trend which is reproduced by the simulations.

Jets with nearby activity have a different jet energy scale, as has been demonstrated
in Monte Carlo simulations~\cite{ATLAS-COM-CONF-2011-060}. 
The systematic uncertainty on their energy scale has been evaluated by studying 
the correlation between the $\pt$ of the tracks associated to the jet and the $\pt$ measured in the 
calorimeter, and contributes to the final uncertainty in the jet energy scale used in this analysis. 

Approximately 40\% of the selected events have more than one vertex in the interaction,
indicating the presence of additional proton-proton interactions.
The vertex multiplicity is low enough that, with a luminous region of several mm and
a vertex reconstruction resolution of a few hundred $\mu$m, the impact of merged vertices on
the analysis is negligible. 
For the instantaneous luminosities considered in this paper, the probability that two hard events 
would occur at the same time is negligible.  However, a soft interaction occurring 
in parallel with the hard interaction can produce a contamination of energy from a nearby soft jet.
The average effect of these overlapping interactions on the jet energy scale is accounted for by
an offset correction, and the systematic uncertainty on that correction has
been evaluated\cite{offset-prel}. The impact of this uncertainty on the overall jet energy scale
uncertainty used in this analysis is negligible for the vast majority of events. 
The overlapping interactions can
also impact the jet counting since the resolution of the jet energy reconstruction depends
on the instantaneous luminosity. The effect
becomes small after performing a cut on the fraction of charged particle $\pt$  that originates from 
the event vertex and that is associated to the jet, as described in Section~\ref{sec:event-sel}. 
The Monte Carlo simulation has been shown to describe tracks within jets\cite{Aad:2011gn} and
general features of events with pile-up interactions\cite{offset-prel}. An 
uncertainty due to the efficiency of the cut has been estimated in Section~\ref{sec:unfolding}. 

In summary, the jet energy scale uncertainty is primarily made of three components: the
uncertainty calculated for isolated jets, the uncertainty caused by the presence of nearby
calorimeter deposits, and the flavor composition uncertainty. 
The uncertainty on the
energy scale of isolated jets is the largest contributor to the total uncertainty in most bins, except
for jets in the five and six-jet bins and of $\pt<200\GeV$, for which the flavor composition 
uncertainty is comparable. The positive systematic uncertainty on the jet energy scale of 
isolated jets falling in the barrel and in high-multiplicity bins varies from 5\% at $60\GeV$ to 
2.5\% at $1\TeV$. In the three-jet and four-jet bins, where the flavor composition is better 
constrained, the systematic uncertainty is at most 3.5\%. The negative systematic uncertainty
is smaller and $\approx 3\%$ across all $\pt$ in the barrel. The impact of nearby calorimeter
deposits is small, increasing the overall uncertainty by at most 1\%. 
The uncertainty is propagated
to the measured distributions using the ALPGEN+HERWIG/JIMMY Monte Carlo simulation 
and varying the $\pt$ of all jets in the event up or down according to the estimated uncertainties. 
The use of the same procedure in the data yields comparable results, but the results obtained
in the Monte Carlo simulation are favored to eliminate the impact of statistical uncertainties in 
the data in bins with few events.

\section{Results}
\begin{figure}[htbp]
 \centering
   \includegraphics[width=0.45\textwidth]{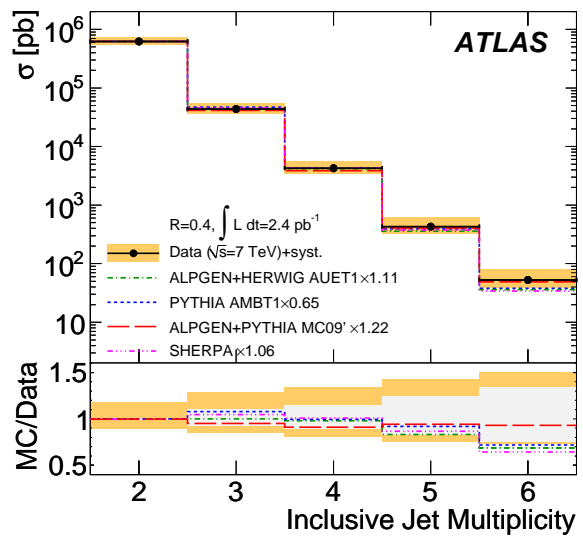}
   \caption{Total inclusive jet cross section as a function of multiplicity.  The data are compared to 
   leading-order Monte Carlo simulations (ALPGEN+HERWIG AUET1, 
   ALPGEN+PYTHIA MC09$^{\prime}$, PYTHIA AMBT1 and SHERPA) normalized to the measured inclusive two-jet cross section.  
   The darker (orange) shaded error bands correspond to the systematic 
   uncertainties on the measurement, excluding the luminosity uncertainty.  The lighter (grey) shaded error band corresponds to the systematic uncertainty
   on the shape of the measured distribution.  
   A plot of the ratio of the different Monte Carlo simulations to the data is presented at the bottom 
   of the 
   figure.  }
   \label{fig:njets}
  \end{figure}

\begin{figure}[htbp]
 \centering
   \includegraphics[width=0.45\textwidth]{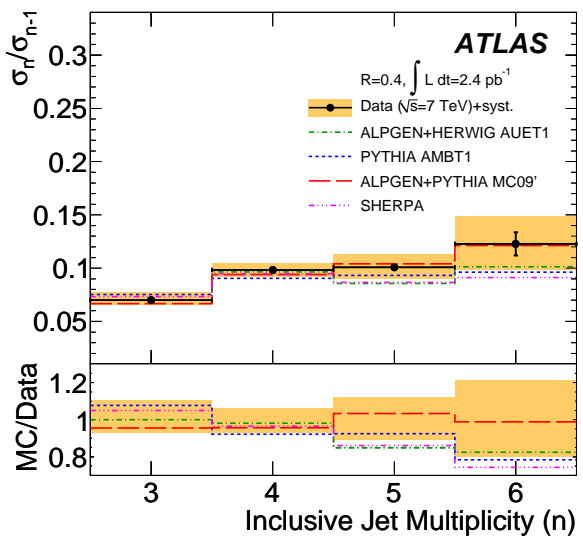}
   \caption{Ratio of the $n$-jet cross section to the $(n-1)$-jet cross section for values of $n$ 
   varying from three to six.  Systematic uncertainties on the 
   cross section ratios are shown as an error band.  
   Other details are as in the caption to Figure~\ref{fig:njets}.} 
   \label{fig:njetsratio}
  \end{figure}

\begin{figure*}[b]
\begin{center}
\subfigure[]{\label{F:diffcrosssec_pt1}
\includegraphics[width=0.4\textwidth]{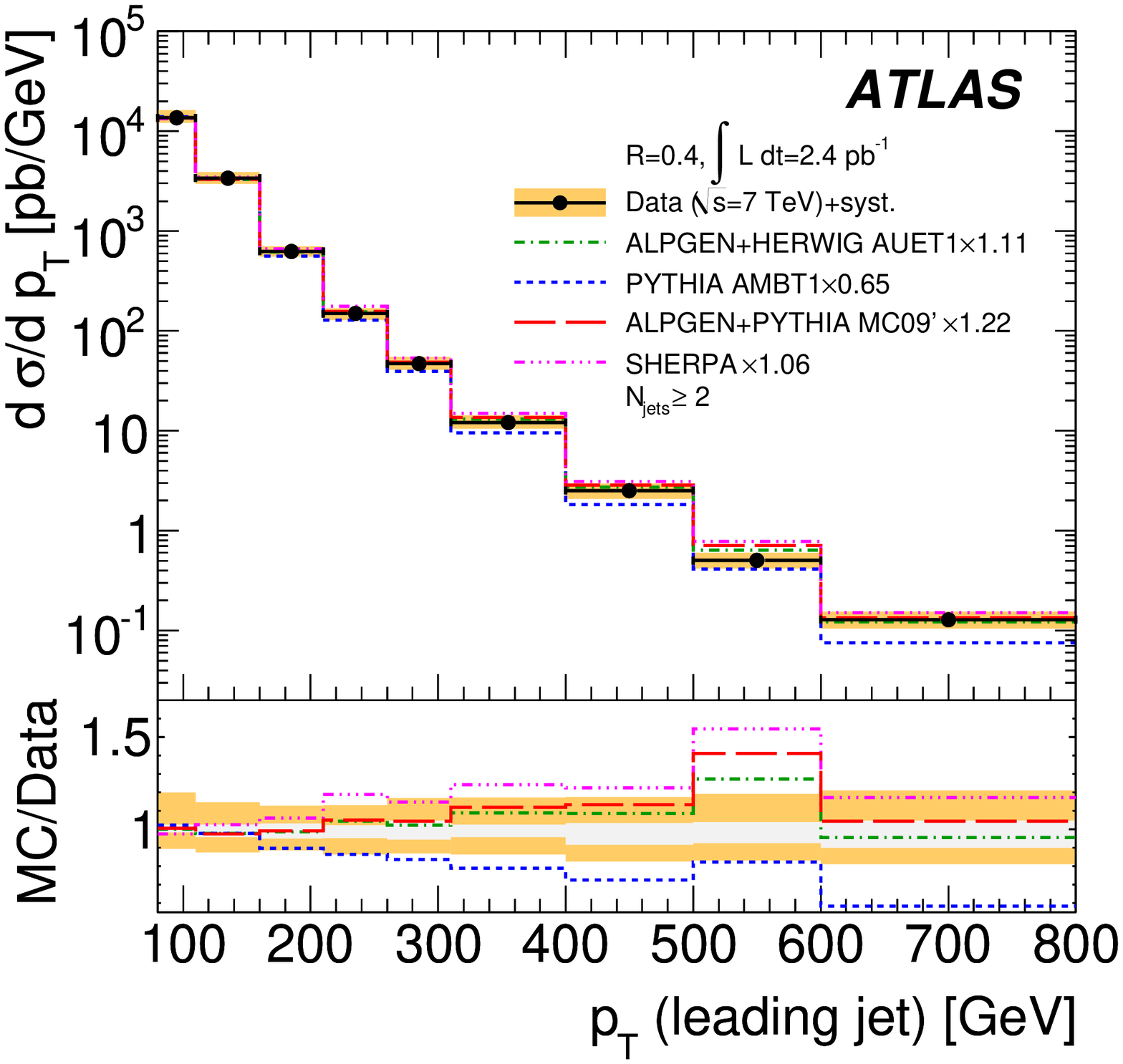} }
\subfigure[]{
\label{F:diffcrosssec_pt2}
\includegraphics[width=0.4\textwidth]{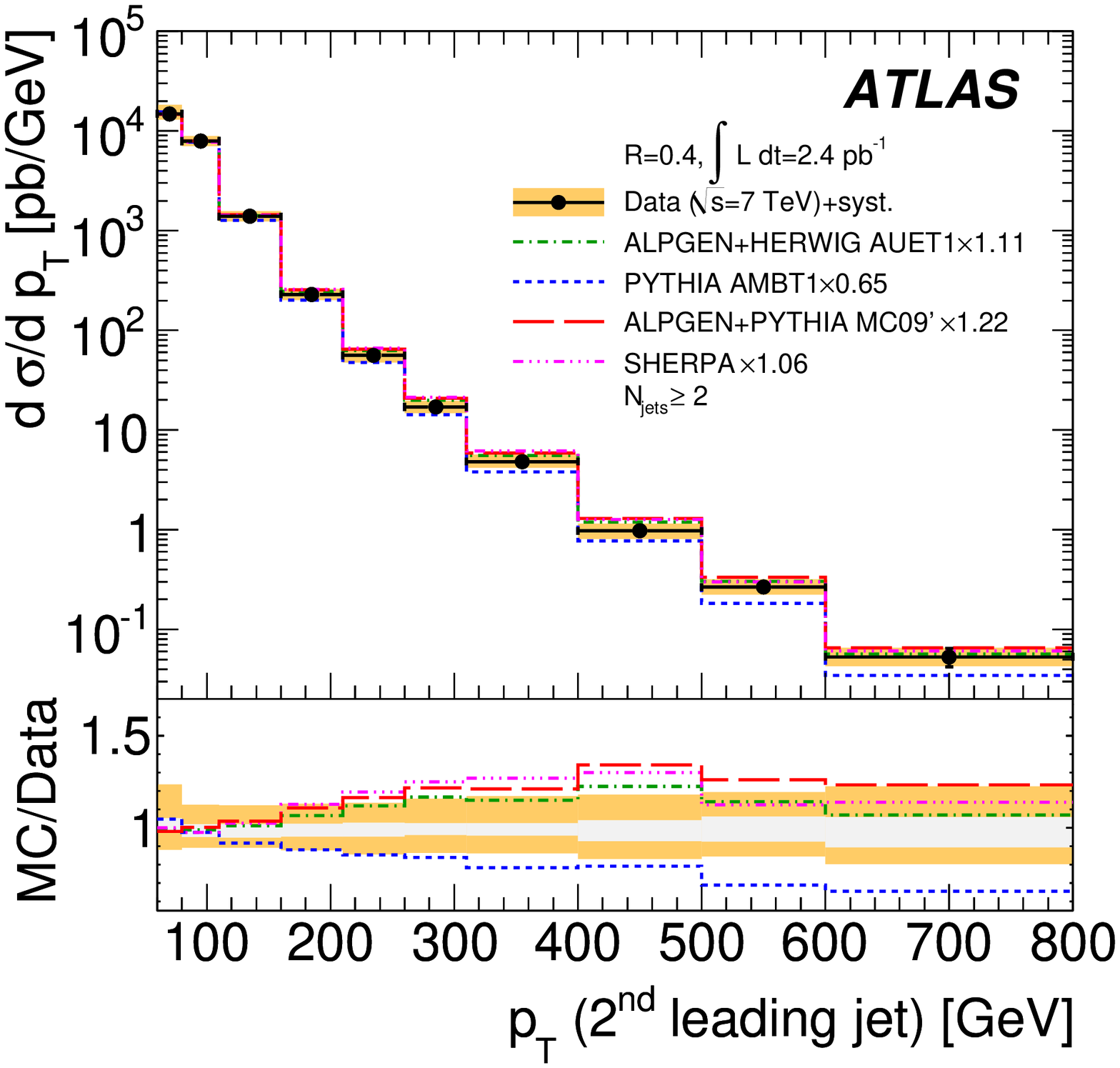} }
\subfigure[]{
\label{F:diffcrosssec_pt3}
\includegraphics[width=0.4\textwidth]{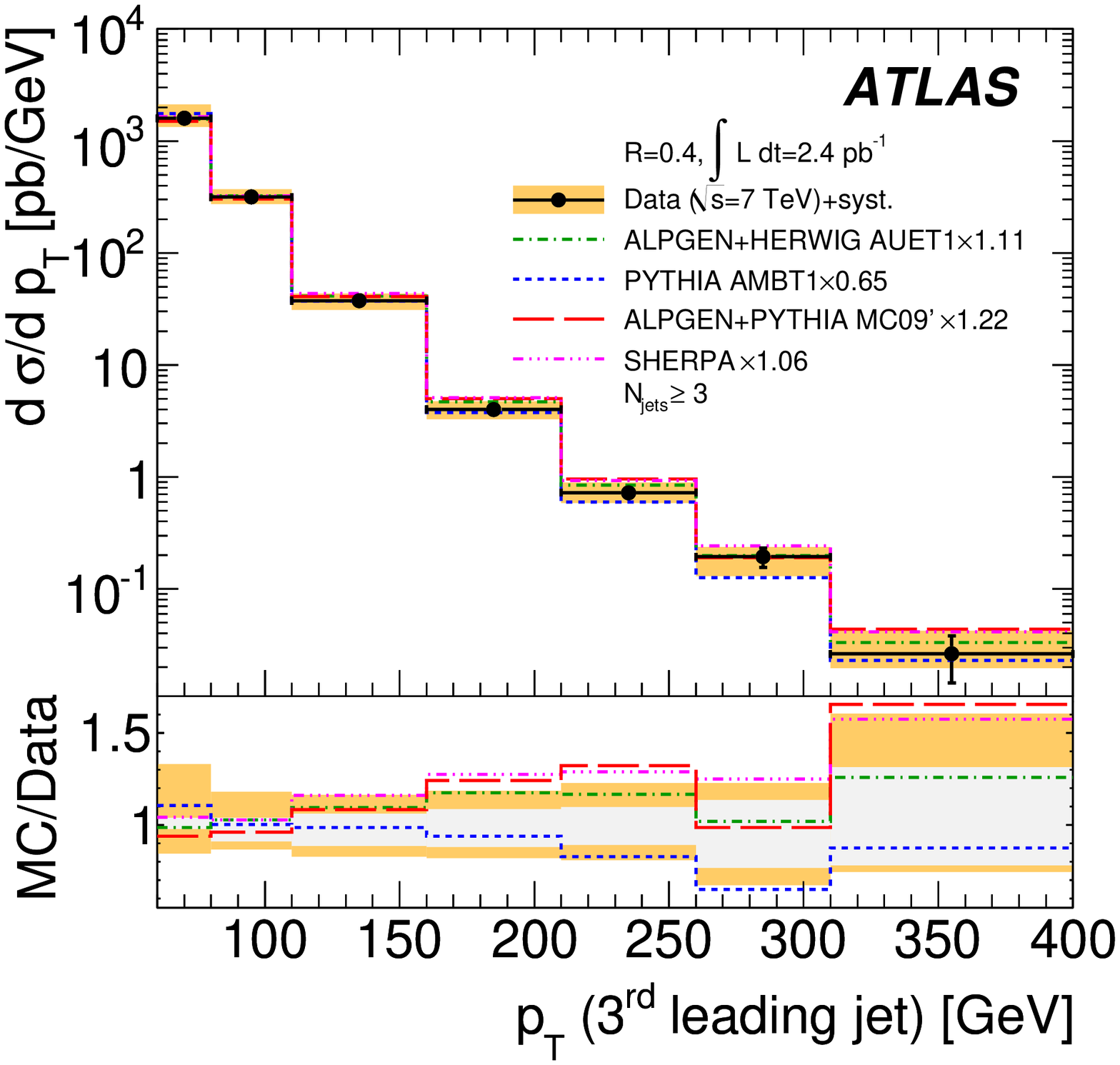} }
\subfigure[]{
\label{F:diffcrosssec_pt4}
\includegraphics[width=0.4\textwidth]{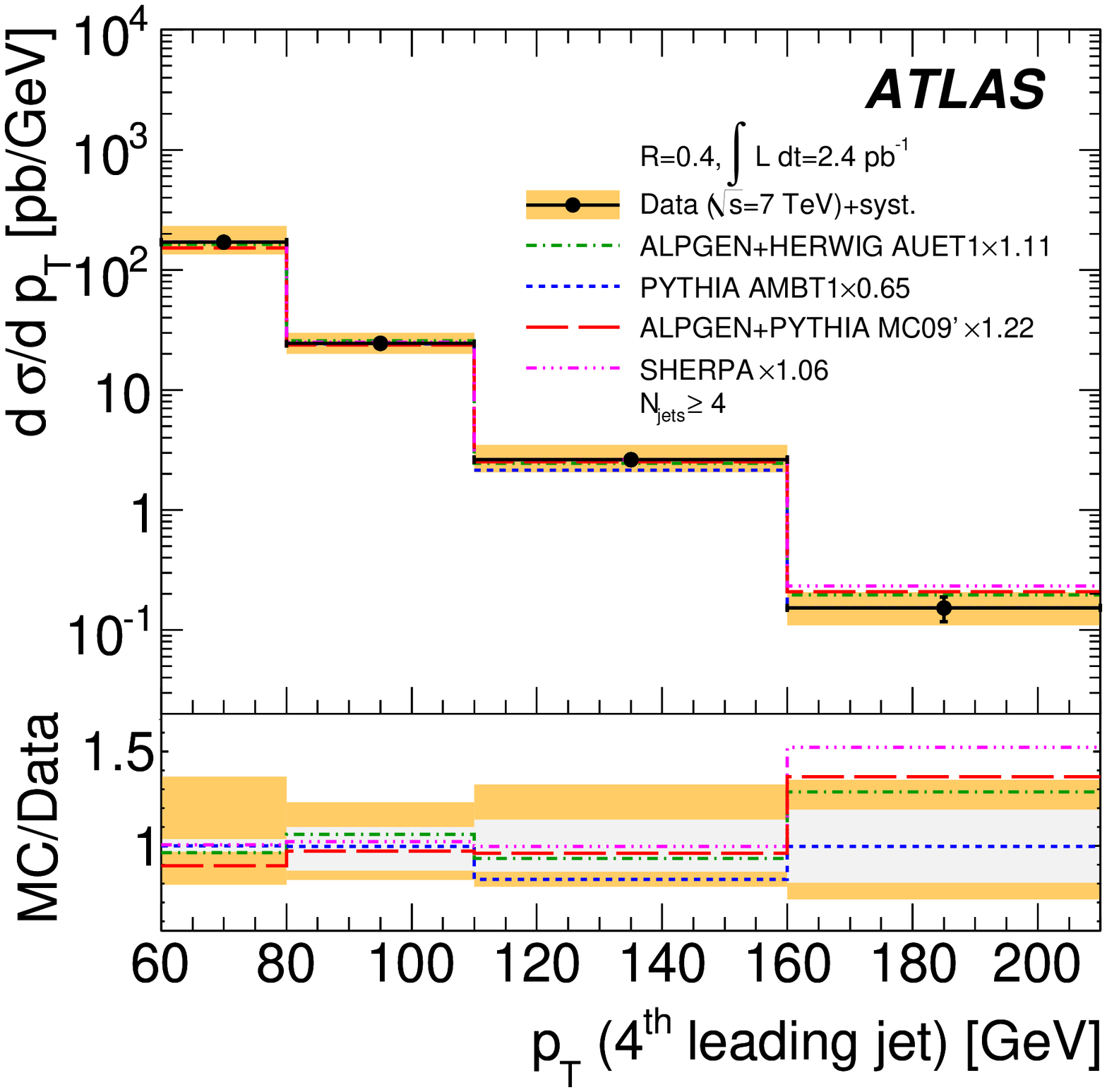} }
\caption{ Differential cross section as a function of  
leading jet $\pt$ for events with $N_{\rm jets}\geq 2$ (a), 
$2^{\rm nd}$ leading jet $\pt$ for events with $N_{\rm jets}\geq 2$ (b), 
$3^{\rm rd}$ leading jet $\pt$ for events with $N_{\rm jets}\geq 3$ (c) and 
$4^{\rm th}$ leading jet $\pt$ for events with $N_{\rm jets}\geq 4$ (d). 
The results are compared to different leading-order Monte Carlo simulations normalized 
to the measured inclusive two-jet cross section.
Other details are as in the caption to Figure~\ref{fig:njets}.} 
\label{fig:diffcrosssec}
\end{center}
\end{figure*}

\begin{figure}[h!tbp]
\begin{center}
\subfigure[]{
\label{F:diffcrosssec_ht2}
\includegraphics[width=0.38\textwidth]{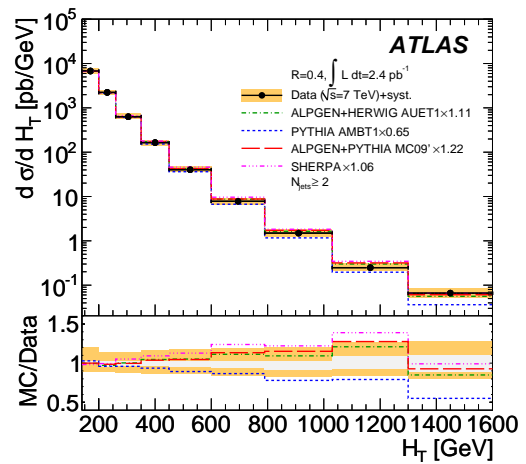} }\\
\subfigure[]{
\label{F:diffcrosssec_ht3}
\includegraphics[width=0.38\textwidth]{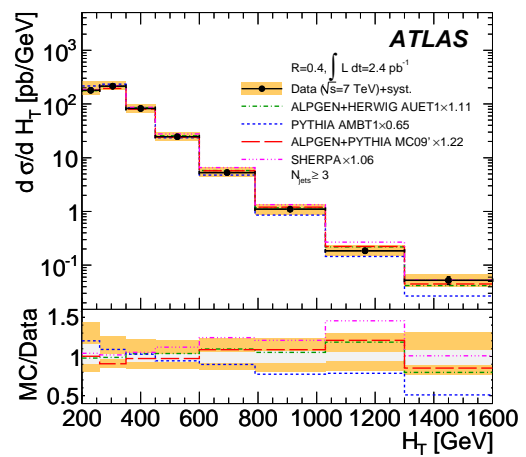} }\\
\subfigure[]{
\label{F:diffcrosssec_ht4}
\includegraphics[width=0.38\textwidth]{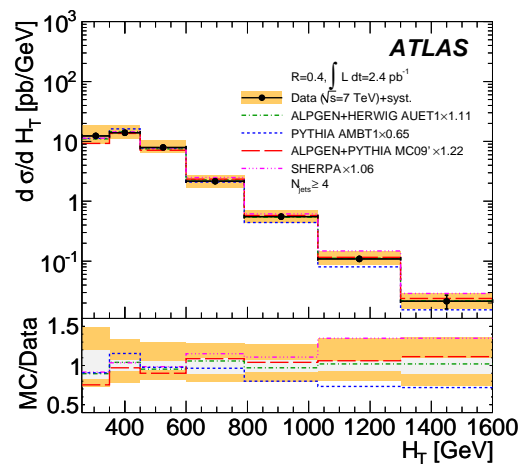} }
\caption{Differential cross section as a function of $\HT$ for events with at least two selected
jets (a), three selected jets (b) and four selected jets (c). 
The results are compared to different 
leading-order Monte Carlo 
simulations normalized to the measured inclusive two-jet cross section. 
Other details are as in the caption to
Figure~\ref{fig:njets}.} 
\label{fig:diffcrossecHT}
\end{center}
\end{figure}

\begin{figure}[htbp]
\begin{center}
\subfigure[]{
\includegraphics[width=0.38\textwidth]{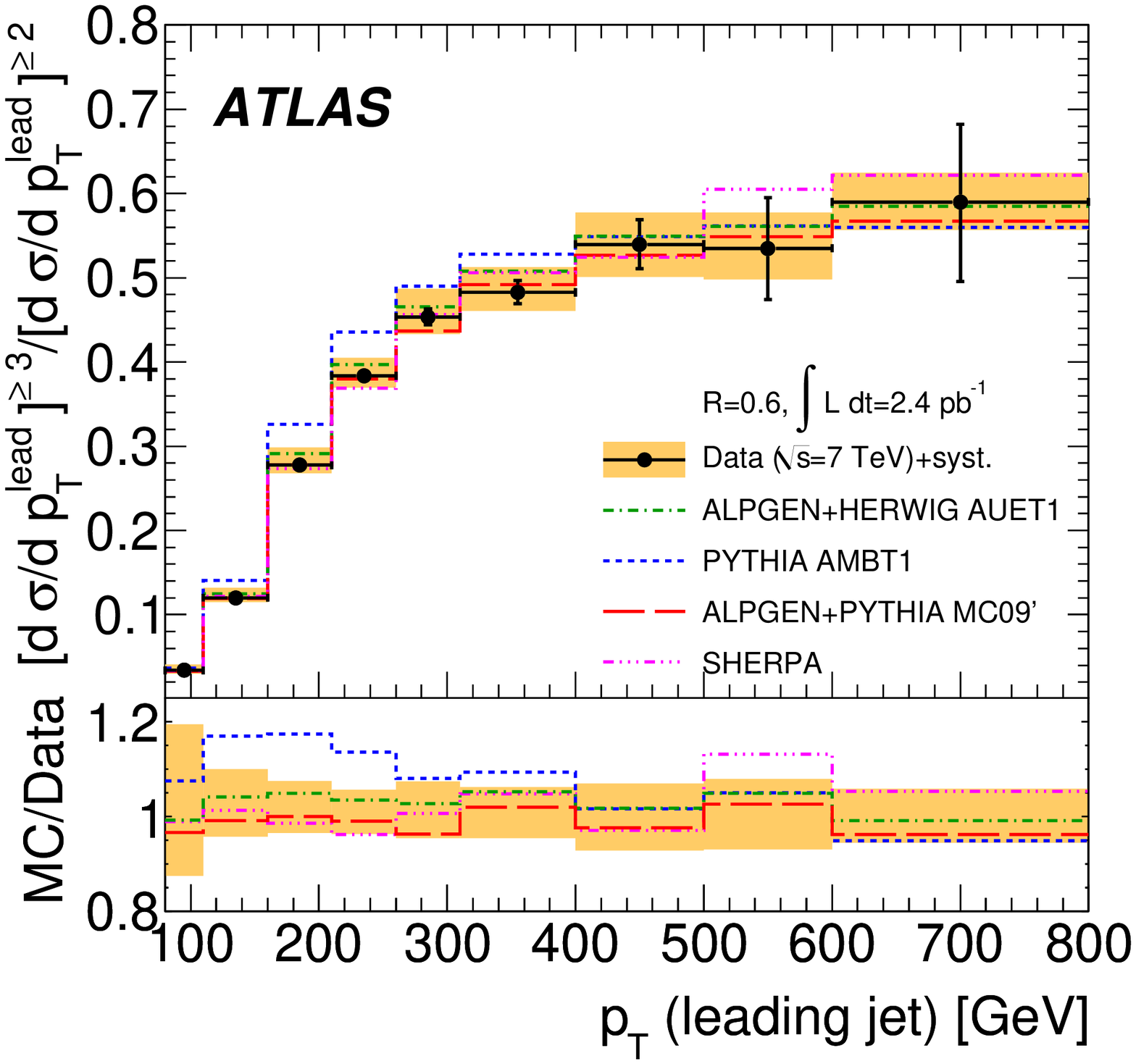} }\\
\subfigure[]{
\includegraphics[width=0.38\textwidth]{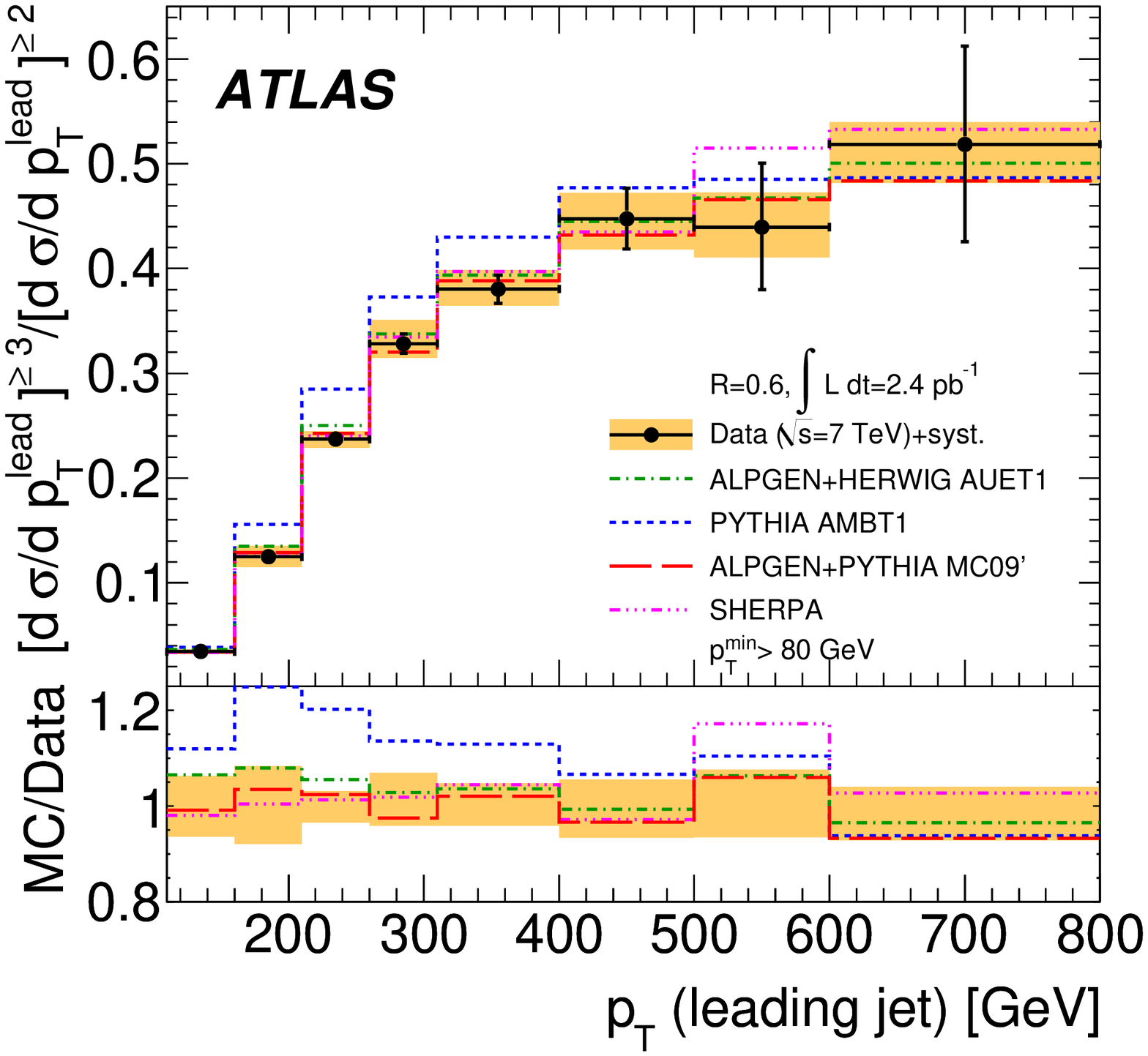} }\\
\subfigure[]{
\includegraphics[width=0.38\textwidth]{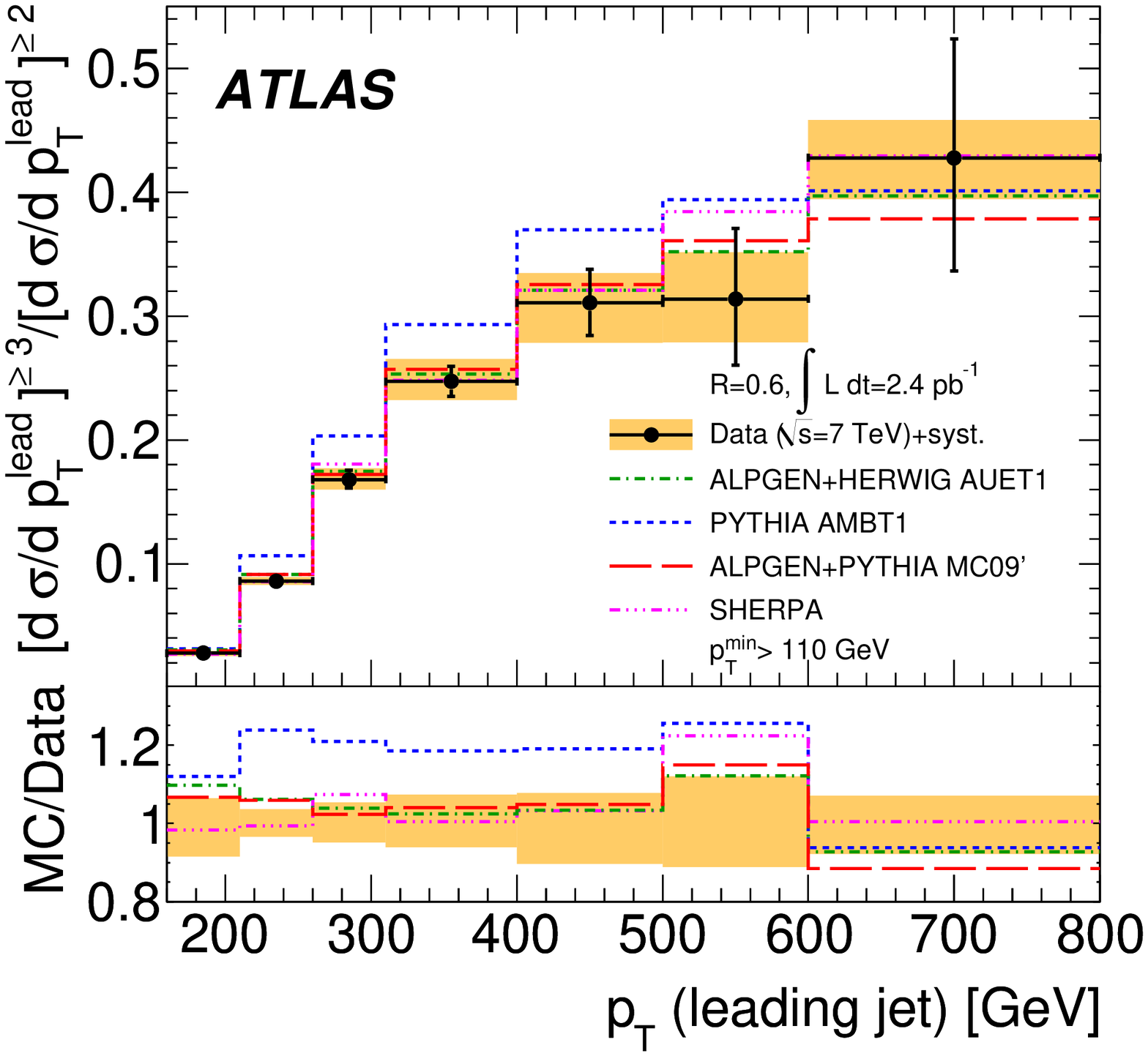} }
\caption{Three-to-two-jet differential cross-section ratio as a function of the leading jet $\pt$.  In the figures, a resolution parameter $R=0.6$ is used.  The three figures contain a minimum $\pt$ cut for all non-leading jets of (a) 60$\GeV$, (b) 80$\GeV$ and (c) 110$\GeV$.
The results are 
compared to leading-order
Monte Carlo simulations. Other details are as in the caption to Figure~\ref{fig:njets}. 
 \label{fig:threetotwoLO_lead}}
\end{center}
\end{figure}

\begin{figure}[!htbp]
 \centering
 \subfigure[]{
   \includegraphics[width=0.38\textwidth]{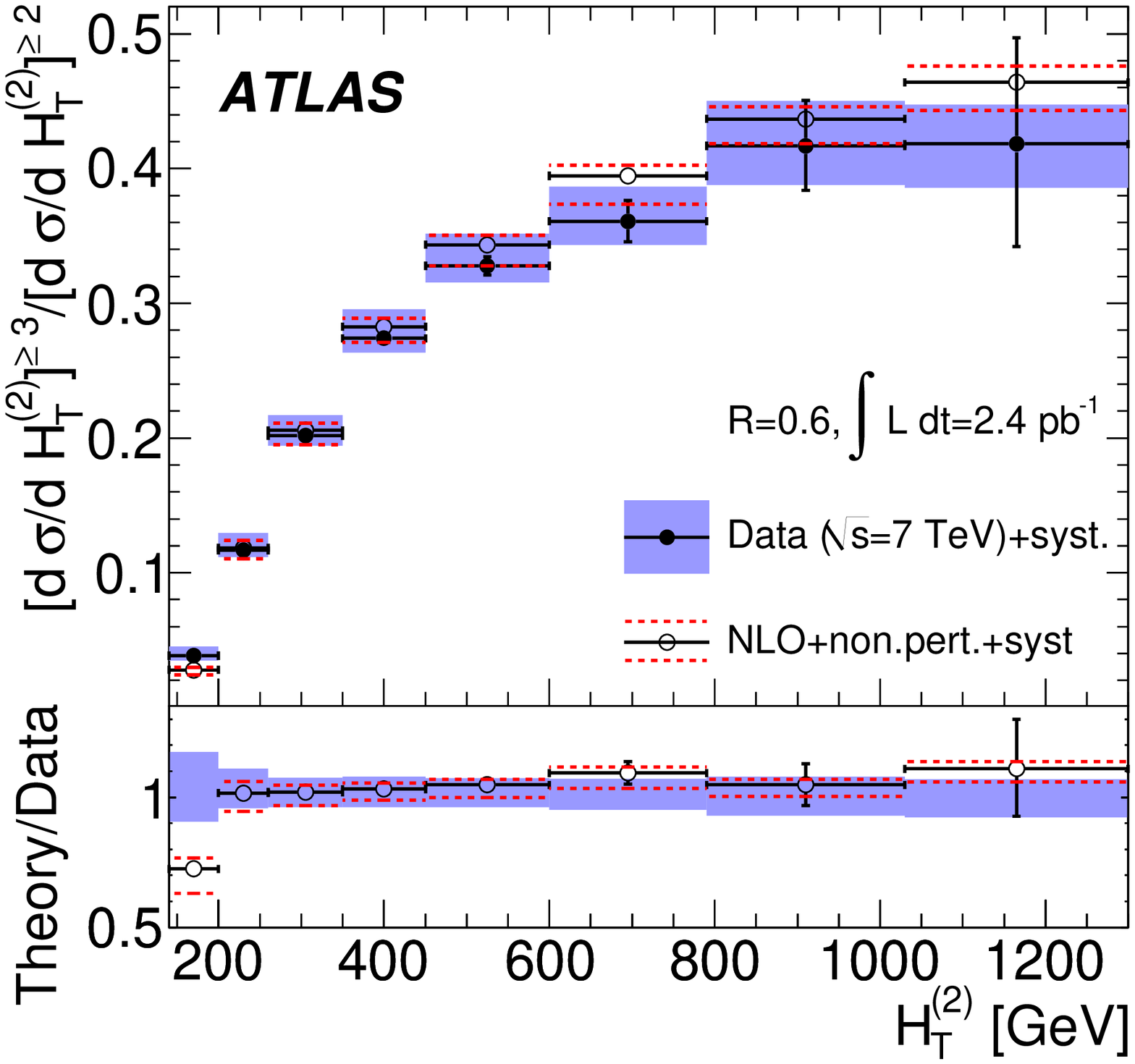}
   }
 \subfigure[]{
   \includegraphics[width=0.38\textwidth]{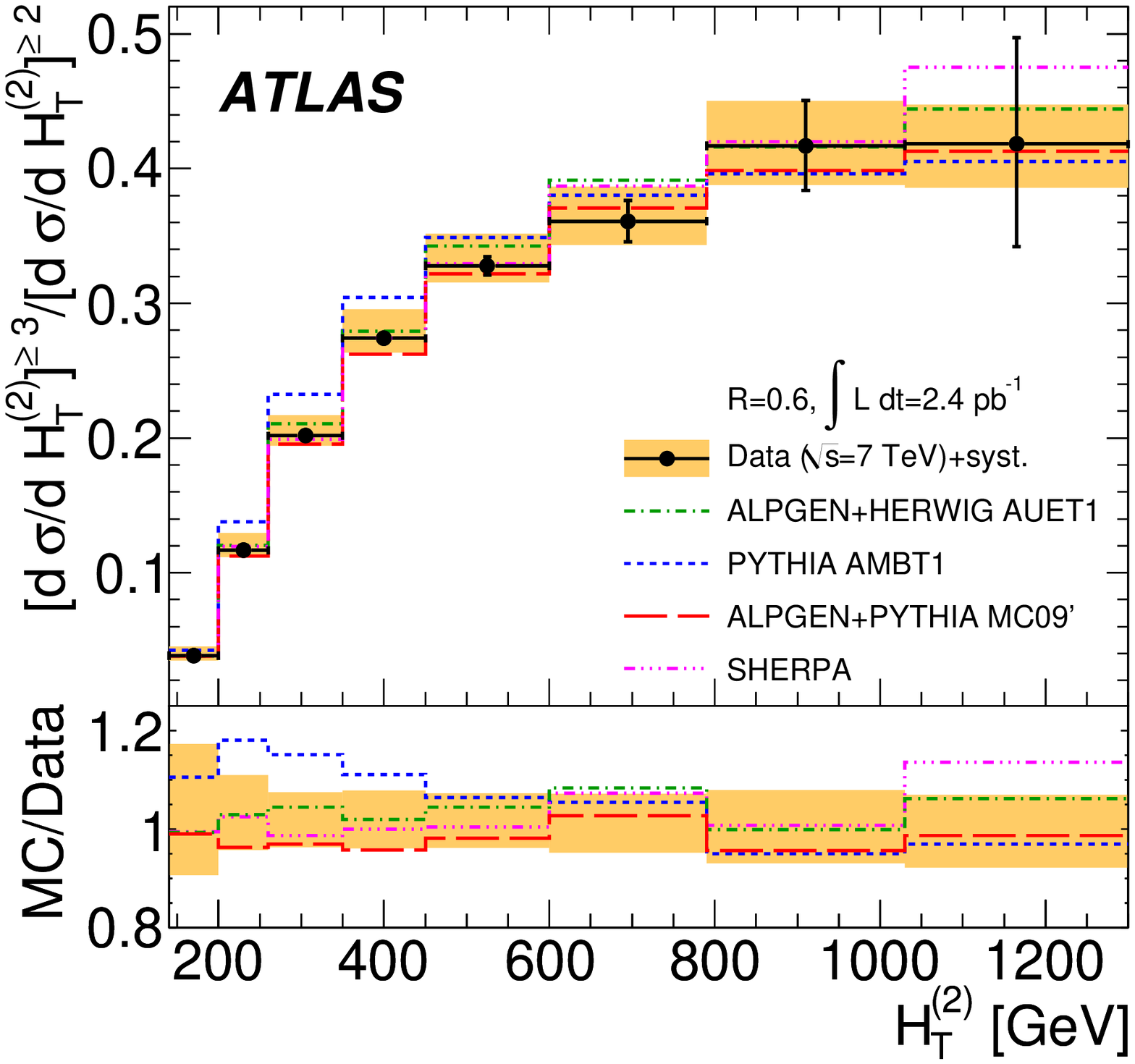}
   }
      \caption{Three-to-two-jet differential cross-section ratio as a function of the sum of the $\pt$ 
   of the two leading jets ($H_T^{(2)}$) using $R=0.6$.  
  The two figures present the same measurements and error bands.
 The data are compared to (a) a NLO pQCD calculation and (b) several leading-order Monte Carlo simulations.
  The systematic uncertainties on the theoretical prediction for the NLO pQCD calculations are 
   shown as dotted red lines above and below the theoretical prediction.  
 }
   \label{fig:threetotwoNLO}
\end{figure}

In this section, measurements\footnote{All measurements in this section have been compiled in tables that can be found in HEPDATA.  The NLO pQCD calculation results are also presented in the tables when applicable.} corrected to the particle level are compared to theoretical predictions. 
For comparisons to leading-order Monte Carlo simulations, the anti-$k_t$ algorithm with resolution parameter $R = 0.4$ 
is used to define a jet.  
In Figures~\ref{fig:njets}-\ref{fig:threetotwoLO_lead} and~\ref{fig:threetotwoNLO}(b), 
the darker (orange) shaded error band bracketing the measured cross section corresponds to the total 
systematic uncertainty, evaluated by adding the individual systematic uncertainties in quadrature 
but excluding the uncertainty coming from the luminosity measurement. 
The ratio of the predictions from the Monte Carlo simulations  
to the measurements is shown at the bottom of each figure.  For Figures~\ref{fig:njets}, 
\ref{fig:diffcrosssec} and~\ref{fig:diffcrossecHT}, the lighter (grey) error band that appears in the ratio of the predictions from the Monte Carlo simulations to the measurements 
represents the total systematic uncertainty on the shape of the measured distributions. 

Only a few representative Monte Carlo simulations that were studied are shown in the figures and tables.
All Monte Carlo simulations are normalized to the measured inclusive two jet cross section.
The normalization factors applied to the Monte Carlo simulations studied are given in Table~\ref{tab:norms},
and distinctive features of some of the Monte Carlo simulations not shown are discussed when relevant.
\begin{table}[htbp]
  \begin{center}
    \begin{tabular}{|c|c|}\hline
	Leading-order Monte Carlo & Normalization factor \\
	\hline
	ALPGEN+HERWIG AUET1 & 1.11 \\
	ALPGEN+PYTHIA MC09$^{\prime}$ & 1.22 \\
	PYTHIA AMBT1 & 0.65 \\
	SHERPA & 1.06 \\
   \hline
    \end{tabular}
    \caption[MC Normalization]{
Normalization factors applied to each of the Monte Carlo simulations in order to match the measured inclusive two-jet 
cross section. 
      \label{tab:norms}}
  \end{center}
\end{table}
Most ALPGEN Monte Carlo simulations predict an inclusive multi-jet cross section 
similar to the measured cross section,
while the PYTHIA Monte Carlo simulation requires scaling factors
which differ the most from unity.
The differences in the normalization factors between ALPGEN+PYTHIA
MC09$^{\prime}$ and ALPGEN+HERWIG/JIMMY AUET1 illustrate differences between PYTHIA 
and HERWIG/JIMMY and their interplay with the matrix-element and parton-shower 
matching implemented in ALPGEN.  
The normalization factor for SHERPA is found to be the closest to unity. 

Figure~\ref{fig:njets} shows the results for the cross section as a function of the inclusive jet multiplicity.  
The measurement systematics are dominated by the jet energy scale uncertainty and range from 10-20\% at 
low multiplicities to almost 30-40\% at high multiplicities. 
The Monte Carlo simulation predictions agree with the measured results across the full inclusive multiplicity
spectrum, even when comparing just to the shape of the distributions.

A study that reduces significantly the impact of systematic uncertainties is the ratio of the $n$-jet 
to $(n-1)$-jet cross section as a function of multiplicity.  In this ratio, the impact of the 
jet energy scale uncertainty is significantly reduced and the uncertainty due to the luminosity 
cancels out. Figure~\ref{fig:njetsratio} presents the results for such a study.  
Both the uncertainties in the data correction for efficiencies and resolutions 
and the jet energy scale contribute comparably to the total
systematic uncertainty, whereas the statistical uncertainties are 
smaller than the systematic uncertainties, and negligible in most bins.
All Monte Carlo simulations are consistent with the measurements at the present precision, 
yet there is a noticeable spread in the predictions. 
Differences at the level of 15\% are observed between 
PYTHIA AMBT1 and ALPGEN+PYTHIA MC09$^{\prime}$ in the first bin. These differences most likely
arise 
from the difference between the pure parton-shower (with $2\rightarrow 2$ matrix elements) 
implemented in PYTHIA and the parton-shower-matched matrix-element calculation (with 
up to $2\rightarrow 6$ matrix elements) implemented in ALPGEN. 
All ALPGEN+PYTHIA tunes studied are comparable in this measurement.

The differential cross section for multi-jet events as a function of the jet $\pt$ 
is useful for characterizing kinematic features.  
The comparison reveals significant differences between the leading order calculations and the measurements.
Figure~\ref{fig:diffcrosssec} presents the $\pt$-dependent differential cross sections for the 
leading, second leading, third leading and fourth leading jet in multi-jet events.  
The systematic uncertainty in the measurement is 10-20\% across $\pt$ and increasing up
to 30\% for the fourth leading jet differential cross section.  The jet energy scale systematic uncertainty remains the dominant uncertainty in the measurement.  However, the uncertainty is less than 10\% (grey shaded error band) for the leading and second leading jet $\pt$ distributions. 

All Monte Carlo simulations
agree reasonably well with the data (orange darker shaded error band).  However, 
the PYTHIA AMBT1 Monte Carlo simulation predicts a somewhat steeper slope compared to the data as a function of the leading jet $\pt$ and the second leading jet $\pt$, whereas the SHERPA and ALPGEN Monte Carlo simulations predict a less steeply falling slope compared to the data.
When using additional tunes and different PDFs,
Monte Carlo simulations using $2\rightarrow2$ matrix element calculations,
in general, make predictions that fall steeper
than what is found in the data, whereas those using $2\rightarrow n$
matrix element calculations predict less steeply falling
spectra.

The differential cross section for multi-jet production as a function of $\HT$ (the scalar sum of the $\pt$ of selected jets in the event) shows similar properties to the differential cross section as a function of $\pt$.
The $\HT$ distributions are typically used for top-quark studies.  
Figure~\ref{fig:diffcrossecHT}  gives the results for the $\HT$-dependent differential cross sections for three different multiplicities compared to the ALPGEN, PYTHIA and SHERPA Monte Carlo simulations.  
Similar conclusions as those reached in the previous figure can be drawn.
 
\begin{figure}[htbp]
\begin{center}
\subfigure[]{
\includegraphics[width=0.38\textwidth]{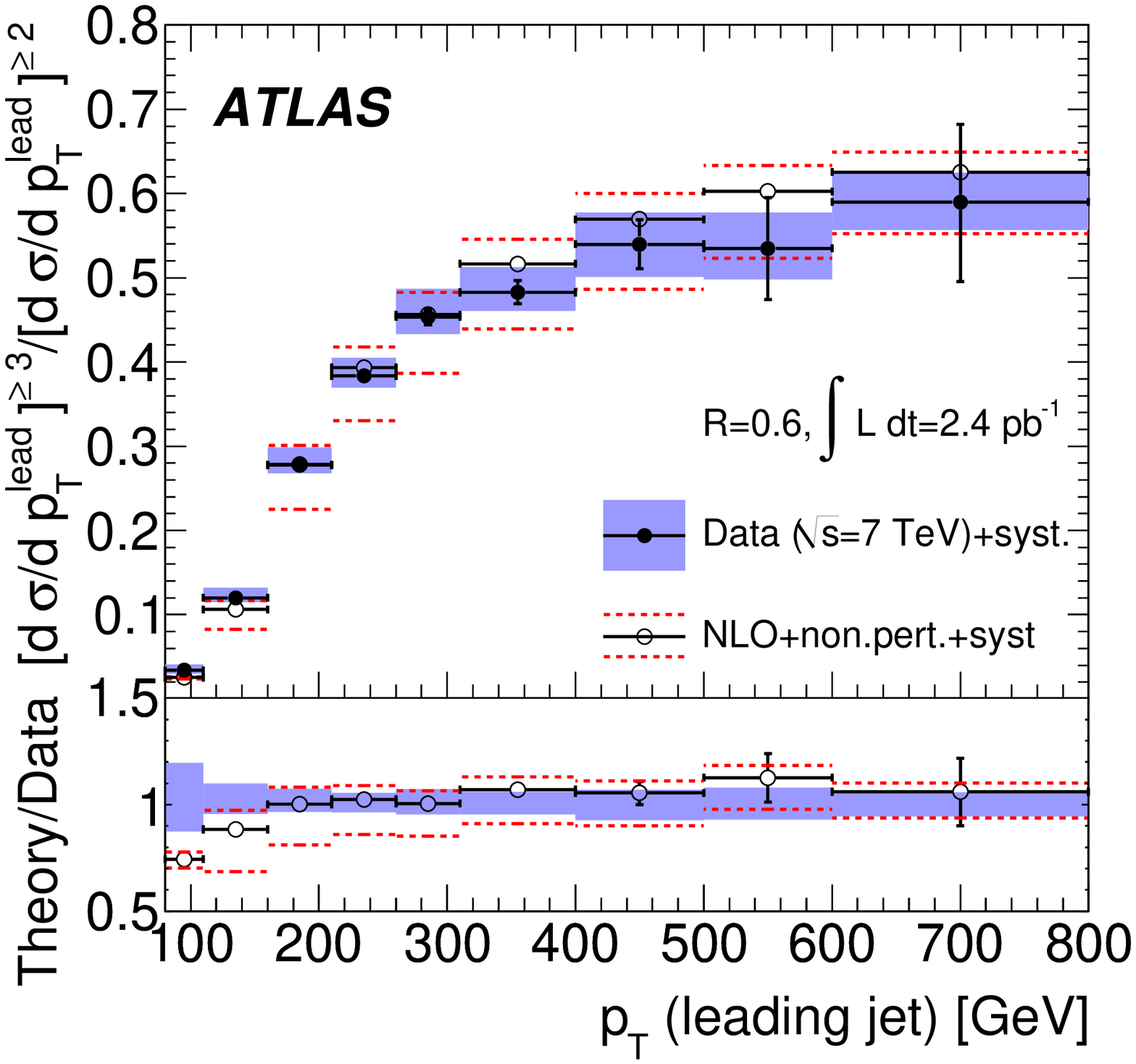} }\\
\subfigure[]{
\includegraphics[width=0.38\textwidth]{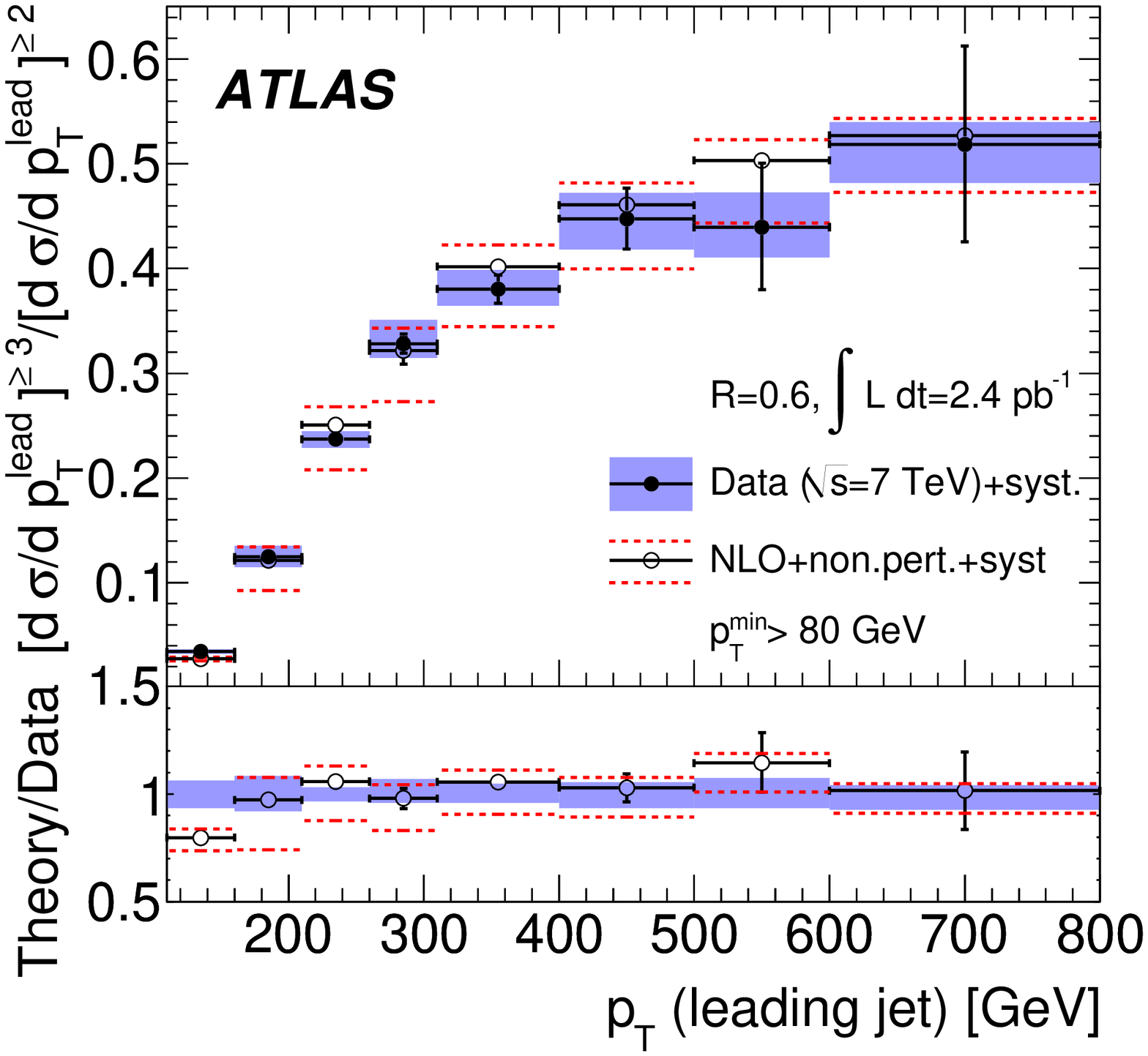} }\\
\subfigure[]{
\includegraphics[width=0.38\textwidth]{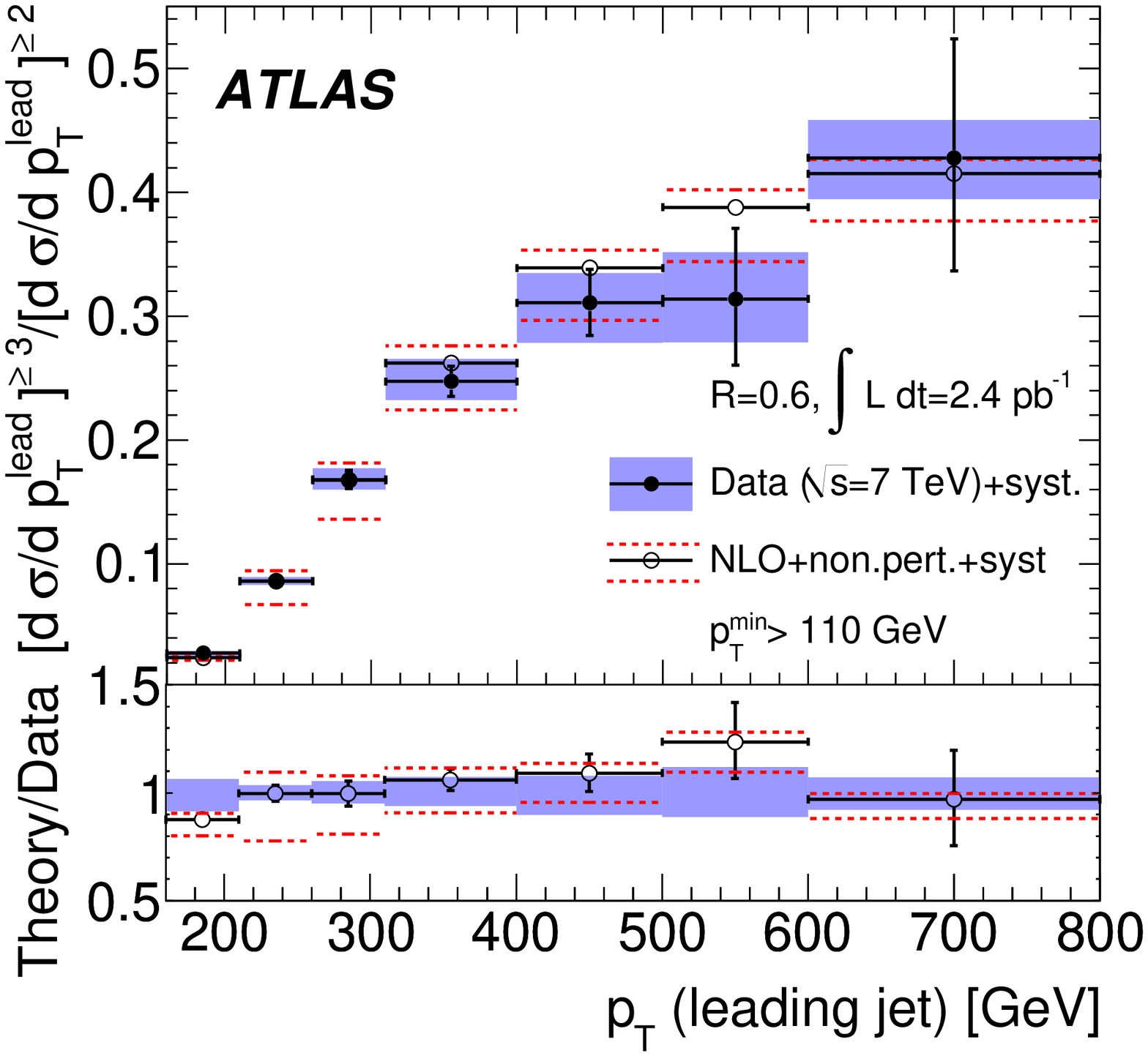} }
\caption{
Three-to-two-jet differential cross-section ratio as a function of the leading jet $\pt$.  
In the figures a resolution parameter $R=0.6$ is used.  The three figures contain a minimum $\pt$ cut for all non-leading jets of (a) 60$\GeV$, (b) 80$\GeV$ and (c) 110$\GeV$.
The results are 
compared to a NLO pQCD calculation with the MSTW 2008 NLO PDF set.
 The data error bands are identical to the results shown in 
 Figure~\ref{fig:threetotwoLO_lead}. 
 The systematic uncertainties on the theoretical prediction are 
 shown as dotted red lines above and below the theoretical prediction. 
  \label{fig:threetotwoNLO_lead}
}
\end{center}
\end{figure}

A measurement with particular sensitivity to limitations in the leading-order Monte Carlo simulations and NLO pQCD calculations is the ratio of the inclusive three-to-two-jet differential cross section as a function of some characteristic scale in the event.  In this measurement, the uncertainty in the luminosity determination cancels out, uncertainties in the jet energy scale are reduced, and statistical uncertainties are limited only by the inclusive three-jet sample.

The three-to-two-jet ratio as a function of the leading jet $\pt$ 
can be used to tune Monte Carlo simulations for effects due to final state radiation. 
Figure~\ref{fig:threetotwoLO_lead} 
presents the results on the measurement of the three-to-two-jet cross section ratio as a function of 
leading jet $\pt$ for jets built with the anti-$k_t$ algorithm using the resolution parameter 
$R=0.6$  and with different minimum $\pt$ cuts for all non-leading jets\footnote{Results (not shown) were also obtained using $R=0.4$ and are compiled in tables in HEPDATA.}. The cut 
on the $\pt$ of the leading jet in the event selection is also increased with the minimum $\pt$
cut ($\pt^{\rm lead}> 110\GeV$ is used in Figure~\ref{fig:threetotwoLO_lead}~(b) and $\pt^{\rm lead}> 160\GeV$ in Figure~\ref{fig:threetotwoLO_lead}~(c)). 
The systematic uncertainties on the measurement are small ($\sim$5\%), except in the lowest $\pt$ bin, where uncertainties in the data correction for efficiencies and resolutions 
and the jet energy scale dominate. 
ALPGEN+HERWIG AUET1 and ALPGEN+PYTHIA MC09$^{\prime}$ describe the data well, and the agreements are largely independent of the tunes chosen.  
SHERPA also describes the data well.  
PYTHIA AMBT1 predicts a higher ratio than that measured over the $\pt$ range from 200$\GeV$ to 600$\GeV$. The disagreement is similar when other 
$2 \rightarrow 2$ 
Monte Carlo simulations with different tunes and PDFs 
are used. 
The systematic uncertainty in the lowest $\pt$ bin decreases significantly as the minimum $\pt$ cut is raised to 80$\GeV$ for all jets. 

Figure~\ref{fig:threetotwoNLO_lead} presents the same measurement results as Figure~\ref{fig:threetotwoLO_lead}, except the data are now compared to the NLO pQCD calculations corrected for non-perturbative effects. The MSTW 2008 NLO PDF set has been used, but comparable results are obtained with the CTEQ 6.6 PDF set. The systematic uncertainties on the theoretical predictions are shown as dotted red lines above and below the theoretical prediction. The NLO pQCD calculations describe the data well, except in the lowest $\pt$ bin, where there is a large discrepancy.  The discrepancy diminishes significantly once the minimum $\pt$ for all jets is raised to 110$\GeV$ and the $\pt$ of the leading jet is required to be greater than 160$\GeV$.  
Additional NLO pQCD calculations of the three-to-two-jet cross section ratio were performed as a function of different kinematic variables, such as $\HT$, the sum of the $\pt$ of the two leading jets ($H_T^{(2)}$) and the sum of the $\pt$ of the three leading jets.  The NLO pQCD calculation for the ratio as a function of $H_T^{(2)}$ was found to give the smallest theoretical scale uncertainty and is, therefore, most sensitive to input parameters such as $\alpha_S$.
Figure~\ref{fig:threetotwoNLO} shows a comparison of the measurement to both 
(a) NLO pQCD and (b) leading order calculations for $R = 0.6$.
 Scale uncertainties of the NLO pQCD calculations
 are larger for jets with $R=0.4$ than with $R=0.6$.
The theoretical uncertainty of the NLO pQCD calculations shown in Figure~\ref{fig:threetotwoNLO}  is comparable to the measurement uncertainties, but is significantly reduced compared to the theoretical uncertainties presented in Figure~\ref{fig:threetotwoNLO_lead} .  With the reduced theoretical uncertainty, the disagreement between data and the NLO pQCD calculations in the lowest $H_T^{(2)}$ bin is now enhanced. Due to the kinematic cuts applied in the analysis, the NLO pQCD calculations only account for the lowest-order contribution to the two-jet cross section in the region where the sum of the first and second leading jet $\pT$ is less than 160$\GeV$.  Consequentially, this effective leading-order estimation is subject to large theoretical uncertainties, which might be responsible for the observed discrepancy.   

A comparison of the same measurement to leading-order Monte Carlo simulations is given in 
Figure~\ref{fig:threetotwoNLO}~(b).  The general agreement between leading-order Monte Carlo simulations with the measurements follows the same general trends as the comparison of the three-to-two-jet ratio versus leading jet $\pt$ shown in Figure~\ref{fig:threetotwoLO_lead}.

\section{Summary and Conclusion}
A first dedicated study of multi-jet events has been performed in proton-proton collisions at a 
center-of-mass energy of $7\TeV$ using the ATLAS detector with an integrated luminosity of 2.4~pb$^{-1}$.  
Leading-order Monte Carlo simulations have been 
compared to multi-jet inclusive and differential 
cross sections. 
The present study extends up to a
multiplicity of six jets, up to jet $\pt$ of $800\GeV$ and up to event $\HT$ 
of 1.6$\TeV$.   

For events containing two 
or more jets with $\pt>60\GeV$, of which at least one has $\pt>80\GeV$, a reasonable agreement is 
found between data and leading-order Monte Carlo simulations with 
parton-shower tunes that describe 
adequately the ATLAS $\sqrt{s}=7\TeV$ underlying-event data. The agreement is found
after the predictions of the Monte Carlo simulations are normalized to the measured
inclusive two-jet cross section. 

All models reproduce the main features of the multijet data.
The $2\rightarrow 2$ calculations 
show some departure from the data for
the three-to-two jet cross-section ratios, predicting a higher ratio than observed.  The $2 \rightarrow n$ calculations describe the measured ratios, independent of the tune or parton shower implementation.  The shape of the differential cross sections as a function of $\pt$ and $\HT$, studied in the inclusive two-jet and three-jet bins,  
falls off less (more) steeply 
in the $2\rightarrow n$ ($2 \rightarrow 2$) calculations.

A measurement of the three-to-two-jet cross section ratio as a function of the leading jet $\pt$ and
the sum of the two leading jet $\pt$s  is described well by 
ALPGEN, SHERPA and a NLO pQCD 
calculation, albeit with a significant discrepancy in the lowest $\pt$ bin for the latter comparison. Future comparisons with 
NLO pQCD calculations will be useful for constraining parameters, such as parton distribution
functions or 
the value of the strong coupling constant, $\alpha_S$. 
Systematic uncertainties from the measurement are presently comparable to the theoretical uncertainties, but should be reduced with larger data samples and higher energy collisions.

\begin{acknowledgement}

We thank CERN for the very successful operation of the LHC, as well as the
support staff from our institutions without whom ATLAS could not be
operated efficiently.

We acknowledge the support of ANPCyT, Argentina; YerPhI, Armenia; ARC,
Australia; BMWF, Austria; ANAS, Azerbaijan; SSTC, Belarus; CNPq and FAPESP,
Brazil; NSERC, NRC and CFI, Canada; CERN; CONICYT, Chile; CAS, MOST and
NSFC, China; COLCIENCIAS, Colombia; MSMT CR, MPO CR and VSC CR, Czech
Republic; DNRF, DNSRC and Lundbeck Foundation, Denmark; ARTEMIS, European
Union; IN2P3-CNRS, CEA-DSM/IRFU, France; GNAS, Georgia; BMBF, DFG, HGF, MPG
and AvH Foundation, Germany; GSRT, Greece; ISF, MINERVA, GIF, DIP and
Benoziyo Center, Israel; INFN, Italy; MEXT and JSPS, Japan; CNRST, Morocco;
FOM and NWO, Netherlands; RCN, Norway; MNiSW, Poland; GRICES and FCT,
Portugal; MERYS (MECTS), Romania; MES of Russia and ROSATOM, Russian
Federation; JINR; MSTD, Serbia; MSSR, Slovakia; ARRS and MVZT, Slovenia;
DST/NRF, South Africa; MICINN, Spain; SRC and Wallenberg Foundation,
Sweden; SER, SNSF and Cantons of Bern and Geneva, Switzerland; NSC, Taiwan;
TAEK, Turkey; STFC, the Royal Society and Leverhulme Trust, United Kingdom;
DOE and NSF, United States of America.

The crucial computing support from all WLCG partners is acknowledged
gratefully, in particular from CERN and the ATLAS Tier-1 facilities at
TRIUMF (Canada), NDGF (Denmark, Norway, Sweden), CC-IN2P3 (France),
KIT/GridKA (Germany), INFN-CNAF (Italy), NL-T1 (Netherlands), PIC (Spain),
ASGC (Taiwan), RAL (UK) and BNL (USA) and in the Tier-2 facilities
worldwide.

\end{acknowledgement}
 \bibliographystyle{atlasnote}
 \bibliography{multijet-paper}
\onecolumn
\clearpage
\begin{flushleft}
{\Large The ATLAS Collaboration}

\bigskip

G.~Aad$^{\rm 48}$,
B.~Abbott$^{\rm 111}$,
J.~Abdallah$^{\rm 11}$,
A.A.~Abdelalim$^{\rm 49}$,
A.~Abdesselam$^{\rm 118}$,
O.~Abdinov$^{\rm 10}$,
B.~Abi$^{\rm 112}$,
M.~Abolins$^{\rm 88}$,
H.~Abramowicz$^{\rm 153}$,
H.~Abreu$^{\rm 115}$
% ...
\bigskip

% ...
$^{10}$ Institute of Physics, Azerbaijan Academy of Sciences, Baku, Azerbaijan\\
$^{11}$ Institut de F\'isica d'Altes Energies and Universitat Aut\`onoma  de Barcelona and ICREA, Barcelona, Spain\\
% ...
$^{48}$ Fakult\"{a}t f\"{u}r Mathematik und Physik, Albert-Ludwigs-Universit\"{a}t, Freiburg i.Br., Germany\\
$^{49}$ Section de Physique, Universit\'e de Gen\`eve, Geneva, Switzerland\\
% ...
$^{88}$ Department of Physics and Astronomy, Michigan State University, East Lansing MI, United States of America\\
% ...
$^{111}$ Homer L. Dodge Department of Physics and Astronomy, University of Oklahoma, Norman OK, United States of America\\
$^{112}$ Department of Physics, Oklahoma State University, Stillwater OK, United States of America\\
% ...
$^{115}$ LAL, Univ. Paris-Sud and CNRS/IN2P3, Orsay, France\\
% ...
$^{118}$ Department of Physics, Oxford University, Oxford, United Kingdom\\
% ...
$^{153}$ Raymond and Beverly Sackler School of Physics and Astronomy, Tel Aviv University, Tel Aviv, Israel\\
% ...
\end{flushleft}


\end{document}